# Python-Fortran Hybrid Programming to Fuse AI and Physical Models: Examples of AI-LDA in climate and weather models (Hf2pMDA_v1.0)

Xianrui Zhu[1+], Zikuan Lin[2+], Shaoqing Zhang[2*], Zebin Lu[2], Songhua Wu[3,4], Xiangyun Hou[2], Zhisheng Xiao[3], Zhicheng Ren[3], Jiangyu Li[5], Jing Xu[5], Yang Gao[6], Rixu Hao[7], Xiaolin Yu[2], Mingkui Li[2], Guangliang Liu[8]

[1]Chongben Honors College, Ocean University of China, Qingdao, 266100, China

[2]Key Laboratory of Physical Oceanography, Ministry of Education, and Institute for Advanced Ocean Study, and Frontiers Science Center for Deep Ocean Multispheres and Earth System (FDOMES), College of Oceanic and Atmospheric Sciences, Ocean University of China, Qingdao, 266100, China

[3]Qingdao Leice Transient Technology Co., Ltd., Qingdao, 266100, China

[4]College of Marine Technology, Ocean Remote Sensing Institute, Ocean University of China, Qingdao, China

[5]Qingdao Marine and Meteorological Institute, Qingdao, 266100, China

[6]Key Laboratory of Marine Environmental Science and Ecology, Ministry of Education, Frontiers Science Center for Deep Ocean Multispheres and Earth System (FDOMES), Ocean University of China, Qingdao, 266100, China

[7]College of Intelligent Systems Science and Engineering, and Engineering Research Center of Navigation Instruments, Ministry of Education, Harbin Engineering University, Harbin, 150001, China.

[8]State Key Laboratory of Physical Oceanography and Artificial Intelligence, Jinan, China.

[+]Co-first authors who contributed equally to this work.

*Correspondence to: Shaoqing Zhang (szhang@ouc.edu.cn)

**Abstract.** Artificial intelligence (AI) provides an unprecedented opportunity for advancing physics numerical modeling including data assimilation, which is a highly efficient and critically-important tool for advancing our understanding on Earth system and its applications. At the same time, deep incorporation of AI and physical modeling can make great driving to advance AI by injecting it rich physics from long time physics-based modeling development. However, since such physics models are conventionally coded in Fortran and AI algorithms usually are conveniently designed in Python, difficulties exist to directly incorporate AI algorithms into physics models, vice versa. Here, based on the F2PY (Fortran to Python interface generator) protocol, we have developed a procedure that implements an infrastructure which conveniently conducts Python and Fortran hybrid modeling and data assimilation ($H_{f2p}$MDA) to form a program entity so that AI algorithms and physical models can invoke mutually. As examples, within $H_{f2p}$MDA, a climate coupled data assimilation (CDA) system is naturally upgraded to a strongly CDA (SCDA) system, and a 1 km high-resolution weather DA system is conveniently implemented within a multi-layer downscaling model that has multiscale DA in different nesting layers. In the climate SCDA system, a coupled general circulation model (CGCM) and a multiscale filtering algorithm is integrated by a Python main controller (PMC) that calls Fortran CGCM components and Weakly-CDA modules as well as a data-trained SCDA algorithm by latent space autoencoder (AE) in Python. In the high-resolution weather DA system, the downscaled model consisting of traditional Fortran DA modules in all mother domains and Python AE DA algorithm in the central child domain is integrated by a PMC that organizes these components. With convenient realization of deep incorporation of any AI algorithm and physics model, the $H_{f2p}$MDA has a great potential to make progress on both AI and scientific modeling.

## 1 Introduction

Since the first Electronic Numerical Integrator and Computer (ENIAC) was born in the 1940s, numerical weather prediction (NWP) models have been developed progressively (Benjamin et al., 2019) and advanced rapidly in recent years (Haarsma et al., 2016). An NWP model is a set of discretized momentum (dynamics) and energy (thermodynamics) budget equations of multi-sphere fluid motions, which can be dated back to the 17th century Newtonian mechanics (Newton, 1687) and 19th century Mayer's and Joule's energy conservation law (Mayer, 1842; Joule, 1843). Nowadays, an NWP model has advanced as an Earth system model (ESM) that consists of coupled atmosphere, ocean, sea-ice, land as well as biogeochemical processes etc. multi-sphere components (e.g., Kay et al., 2015; Danabasoglu et al., 2020). Now, the ESM is pursuing precise simulation and prediction for weather-climate variations through resolving multiscale interactions in the geofluid by developing high-resolution (HR) model (Zhang et al., 2023; Mouallem et al., 2025), which is a critically-important platform for Earth science studies (Haarsma et al., 2016).

However, as the spatial resolution continuously increases, physics-based modeling encounters many challenges (Chang et al., 2020; Marotzke, 2023). For example, as a consequence of water phase changing with strong nonlinearity, the genesis and melting processes of sea-ice have scales from millimeters (freezing coagulations, for instance) to hundreds of kilometers (glaciers, for instance) and rich scale processes such as melt ponds (e.g. Feng et al., 2022), ice leads (e.g. Qu et al., 2024) and polynyas (e.g. Diao et al., 2022) between them. Each scale band plays a relatively independent and important role (Golden et al., 2020). Such a multiscale nature makes particular difficulties on HR modeling of sea-ice, since any HR model needs a specific parameterization for its sub-grid processes (Gou et al., 2025). The other outstanding example on HR ESM challenges is parameterization of planetary boundary layer (PBL) processes (Jia and Zhang, 2020). When the model horizontal resolution comes to a level of kilometers, classic PBL parameterization that deals with low resolution sub-grid boundary processes is no longer suitable for describing turbulences induced by differently-featured underneath underlying surface structures, such as mechanical (roughness), thermal (urban heat) and shear-induced turbulences etc. (Kadivar et al., 2021).

Machine learning artificial intelligence (AI) technology provides unprecedented opportunities to resolve the challenging issues of physics-based HR modeling described above. As long as data sufficiently represent an end-to-end process, no matter how complex the process is, the process can be resolved by a data-training procedure based on neural (deep learning) network (Hopfield, 1982; Rumelhart et al., 1986). Starting from the basic principle of least-square fitting to resolve optimal weighting coefficients on a network, AI algorithms advance very rapidly and help greatly enhance accuracy of weather forecasts (Bi et al., 2023) and climate predictions (Ham et al., 2019). Studies have revealed that applications of AI techniques can help improve traditional physics-based models on frontier challenging issues such as subgrid parameterizations (e.g. O' Gorman & Dwyer, 2018; Han et al., 2020; Yuval & O' Gorman, 2020) and nonlinear data assimilation (e.g. Wahle et al., 2015; Ruckstuhl et al., 2021; Lin et al., 2025). While data-driven AI technology rapidly spreads into almost every engineering and scientific discipline, the data-driven nature limits its further advancement and therefore physics-guided AI merges explosively (Faroughi et al., 2024; Yuan and Guo, 2024).

Deep incorporation of AI and physical models (let's call so for wording convenience) can make unbelievable opportunities to drive both of them even more rapidly advancing by injecting AI rich physics from long time physics-based modeling development and letting physical models use any trained AI algorithm. Nevertheless, physical models are conventionally coded in Fortran (or Fortran-C hybrid) language and AI algorithms are designed and trained in Python language. Although great efforts are made for combining utilities of different languages together (Müller et al., 2025), deep incorporation of AI and physical models is still very difficult because of the lack of convenient infrastructure friendly to physical modelers between Fortran and Python. At present, on the one hand, most of the physics-guided AI employs some simplified physical expressions by Python (e.g. Häfner et al., 2018; Kochkov et al., 2024; Shu et al., 2025; Hao et al., 2025). On the other hand, physical models can only use some selected AI algorithms that can be translated into Fortran (Heuer et al., 2024) with FTorch library that does still not fully support all Torch application interfaces (Atkinson et al., 2025).

In this study, we document our developing process of Python-Fortran hybrid programming to establish a convenient infrastructure platform on which AI and physical modeling and data assimilation can deeply incorporate with respect to each other, called $H_{f2p}$MDA, based on the F2PY (Fortran to Python interface generator) protocol (Harris et al., 2020). Basically, through 3 steps of preparations in environment setting, recompiling of Fortran codes of physical model and plug interface, and simple programming of Python main controller, $H_{f2p}$MDA conveniently realizes the online deep incorporation of AI and physical modeling and data assimilation. Through exhibitions of two applications on a coupled climate model and a regional weather model, we show that $H_{f2p}$MDA is a readily user-friendly platform to physical modelers and researchers. The former presents an application of AI latent space data assimilation (LDA) to implement strongly-coupled DA which has long been challenging in climate studies, while the latter confirms that the AI-LDA can render high-precision weather analysis by assimilating Lidar measurements which include turbulence information.

This paper is organized as follows. After the introduction, **Sect. 2** describes the baseline of F2PY protocol and outlines the general procedure to construct the framework of online Python-Fortran hybrid modeling and data assimilation ($H_{f2p}$MDA) and as an example of AI algorithms applied in $H_{f2p}$MDA, **Sect. 3** describes a variational autoencoder-based AI-LDA algorithm.

**Sections 4 and 5** give examples of detailed implementation of applying $H_{f2p}$MDA to a climate model to realize strongly-coupled data assimilation and a weather model to fulfill fine-scale convection data assimilation with Lidar measurements. Finally, summary and discussions are given in **Sect. 6**.

## 2 $H_{f2p}$MDA - F2PY-based Python-Fortran hybrid programming to fuse AI and physical model and data assimilation

### 2.1 Implementation of $H_{f2p}$MDA based on the F2PY protocol

The F2PY protocol (Harris et al., 2020) provides a basic idea for designing online Python-Fortran hybrid computation. A Python coded program with a kind of interpreted fashion is directly executed without pre-compiling requirement as compiled languages such as Fortran etc.  A natural way to realize hybrid computation of Python and Fortran is that Python-callable Fortran (PCF) modules can be imported into a Python main controller (PMC) as callable objects and scientific Fortran computation can call Python-coded AI algorithms as subroutine or function. To that end, there are four aspects of concerns that need to be addressed in an infrastructure of Python-Fortran hybrid modeling and data assimilation (Hf2pMDA) designed in this study: (1) computational environment needs to be set to support both the Python and Fortran computations; (2) Fortran model codes need to be recompiled so that they become PCF modules; (3) some of Python codes need to be Fortran-callable so that they become Fortran-callable Python (FCP) modules for which the Fortran program can directly call; (4) a PMC needs to be designed and application-oriented interfaces need to be designed to link PCF modules to PMC, and let scientific Fortran computation being able to call Python-coded AI algorithms. We describe the details of implementation of 4 aspects above as 4 steps in **Text S1: General procedure of implementation of $H_{f2p}$MDA.** Following the procedure above and taking care of some key points noted in **Section 2.2**, we can construct $H_{f2p}$MDA. By these 4 steps, within the $H_{f2p}$MDA, the scientific modeling and machine learning can be fused together as illustrated in **Fig. 1**. On the one hand, it readily realizes "AI for science," i.e., when AI algorithms can help resolve challenging issues in physical modeling and data assimilation, such algorithms can be directly applied to legacy Fortran codes from long-time scientific advancement. On the other hand, the $H_{f2p}$MDA can help on "science for AI," i.e., if necessary, the scientific computation results can directly join the training process. Examples include reinforcement learning in which dynamics is used as a part of training constraint (Banerjee et al., 2025; Sun et al., 2025).

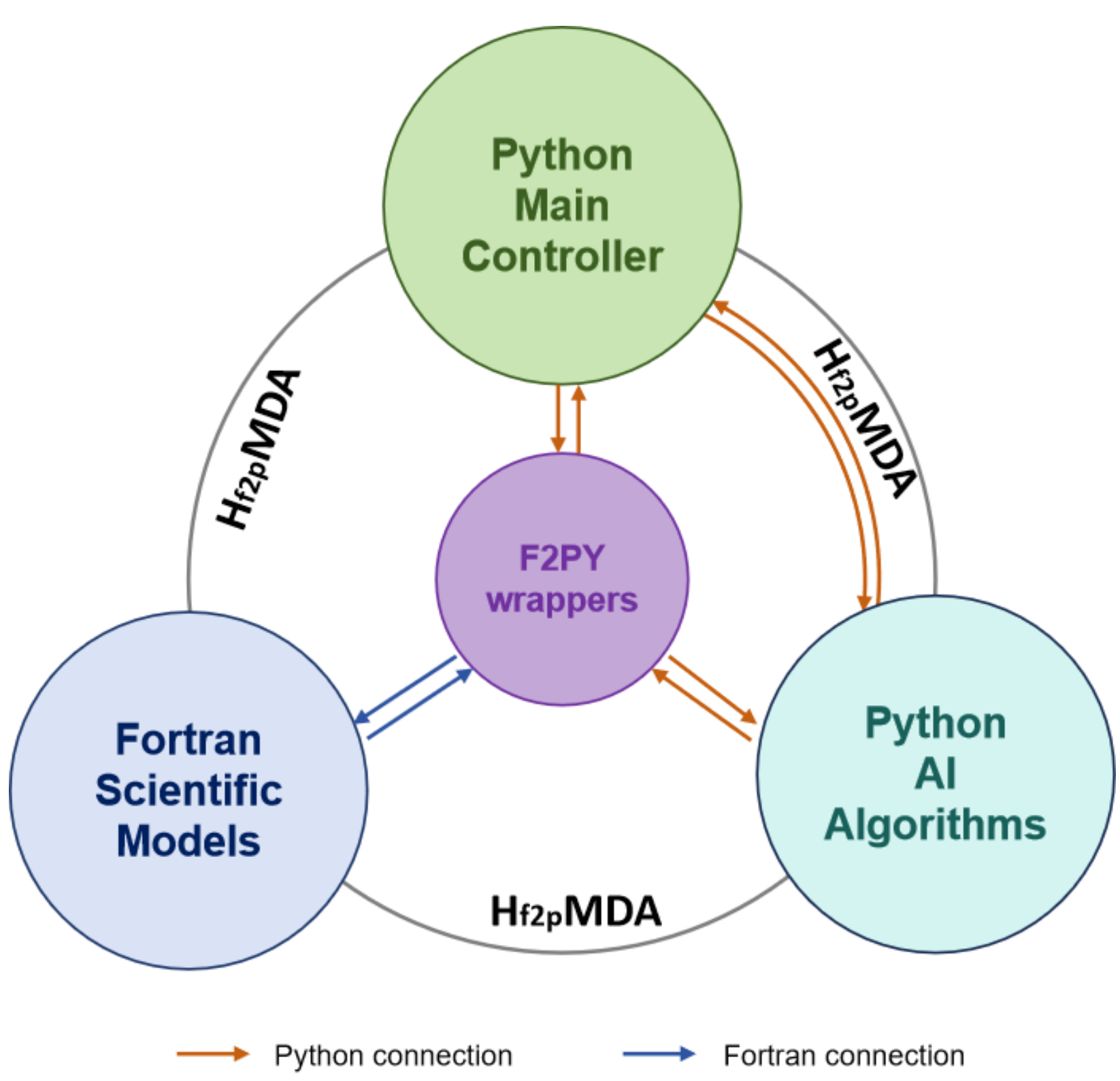

**Figure 1: A schematic illustration of the designed $H_{f2p}$MDA to fuse Fortran scientific models and Python machine learning.**

Next, after addressing innovative key points in Hf2pMDA implementation and Hf2pMDA's merits, we will give application examples of $H_{f2p}$MDA, in which an AI DA algorithm is fused into climate and weather models to fulfill strongly-coupled DA and fine-scale convection DA with Lidar measurements.

### 2.2 Notes of key points in $H_{f2p}$MDA implementation and $H_{f2p}$MDA's merits

The F2PY is a standard utility distributed with NumPy, which is one of the foundational libraries for numerical computing in Python (Harris et al., 2020). It has been widely used to generate Python interfaces for Fortran routines and has become part of the scientific Python build ecosystem, including applications in SciPy (Virtanen et al., 2020), WRF-Python (Ladwig, 2017), GeoCAT-f2py (UCAR/NCAR, 2020), and related tools. Although F2PY has traditionally been used to wrap individual Fortran functions or libraries for Python-side invocation, its potential for coupling legacy scientific models with Python-based algorithms has not been fully exploited. This is partly because F2PY was originally designed for workflows in which Python acts as the main "glue" language, whereas usually physical Earth system models (including data assimilation) are organized around compiled Fortran drivers, deeply nested physical parameterization routines, and tightly controlled time-stepping procedures. As a result, directly introducing Python-based machine learning or data assimilation algorithms into such models is not always straightforward.

In $H_{f2p}$MDA, F2PY is used in two complementary paths to bridge this gap. Here, the first path is referred to as "PMC organizing PCF modules," where a Python main control layer reorganizes and invokes Python-callable Fortran (PCF) modules. The second path is referred to as "Fortran program calling FCP functions," where the original Fortran execution flow is largely preserved and Python functions are invoked through callback interfaces. Together, these two paths allow $H_{f2p}$MDA to support both Python-driven model orchestration and callback-based embedding of Python algorithms into existing Fortran model workflows. The key notes of these two paths will be described in detail in the following 2 subsections combined with **Fig. 3** together.

### 2.2.1 PMC organizing PCF modules

Similar to the strategy adopted in fv3gfs-wrapper (McGibbon et al., 2021), the path of PMC organizing PCF modules decomposes the original Fortran model workflow into callable components. Key model routines and state variables are first exported from the original scientific model and then wrapped as Python-callable Fortran modules through F2PY utilities. The PMC can subsequently invoke these wrapped components, reorganize the execution logic, and insert Python-side algorithms such as pre-processing, machine learning, data assimilation, and post-processing etc.

The major advantage of this path is its conceptual simplicity. Instead of embedding Python directly inside the original Fortran time-stepping loop, the model components are exposed to Python and then controlled externally by PMC. This makes the Python-Fortran coupling workflow easy to understand and debug, and makes algorithm development more flexible. Moreover, the F2PY signature-file mechanism allows users to explicitly specify which Fortran routines should be exposed. Together with support for static or dynamic libraries, this makes it possible to integrate the wrapping process into an existing system without manually writing C glue-code layers. In this sense, F2PY can automatically generate native Python C/API extension modules that connect Python with the compiled Fortran components.

The implementation procedure can be summarized as follows. First, the original Fortran workflow is decomposed into several callable components, and the required data by Python-side algorithms are identified (❶ in **Fig. 2a**). Second, F2PY is used to generate or customize the signature file (❷ in **Fig. 2a**). Third, the scientific model or selected components are compiled into a linkable library and then wrapped by F2PY (❸ in **Fig. 2a**). Finally, the PMC reorganizes the model logic and integrates Python-side algorithms into the workflow (❹ in **Fig. 2a**).

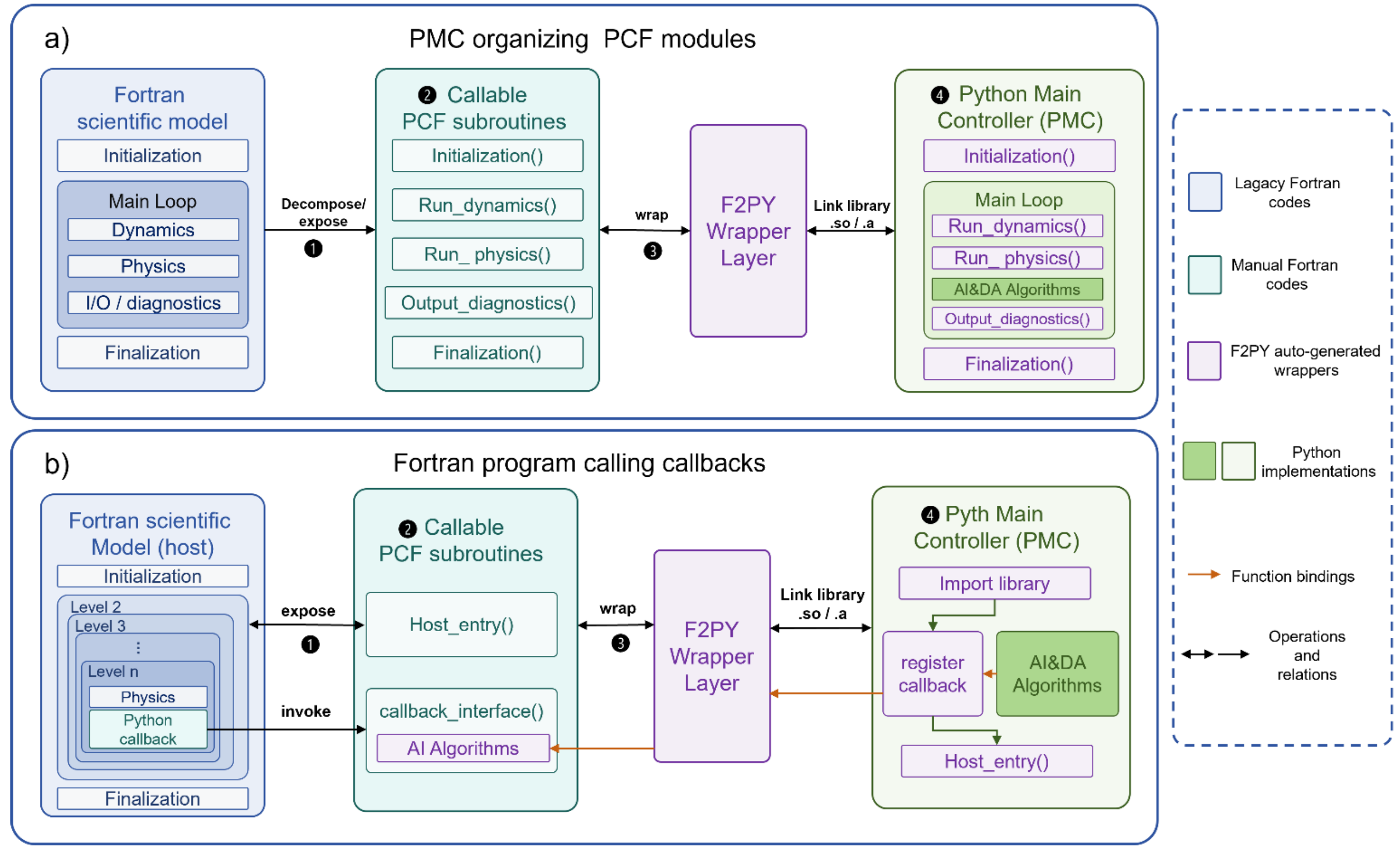

**Figure 2: A schematic illustration of the path of a) PMC organizing PCF modules, and b) Fortran program calling FCP functions in $H_{f2p}$MDA.**

### 2.2.2 Fortran program calling FCP functions

The path of PMC organizing PCF modules is effective when the main model workflow can be reasonably decomposed into independent callable components. However, this strategy can become labor-intensive, or even impractical, for highly undisassembled models (multiple nested layers with recursive routine calls, for instance), especially when the variables required by Python-side algorithms are produced or updated in lower-level routines rather than in the top-level driver. For models such as WRF, whose execution flow involves deeply nested physical and dynamical processes, extracting all relevant components into a Python-controlled workflow may require extensive refactoring and change the original model organization. To resolve this difficulty, $H_{f2p}$MDA also supports a callback-based strategy by the path of Fortran program calling FCP functions. This strategy is conceptually related to previous efforts such as WRF–ML v1.0 (Zhong et al., 2023), Arnold et al. (2024), and Zhang et al. (2025), in which Python-side algorithms are invoked from Fortran model code. The different part of Hf2pMDA is that the original Fortran program remains the primary execution driver, while selected Fortran routines call a callback-facing interface only when Python-side computation is required. The corresponding Python functions are registered through the F2PY-generated layer and are subsequently invoked during model execution.

A key merit of this path is that it reduces intrusive refactoring of the host model. Instead of decomposing the entire model workflow and reconstructing it under Python control, only a thin callback-facing Fortran interface is inserted at the locations where Python-side algorithms need to interact with the host model. Lower-level Fortran routines can pass model states or diagnostic variables to this interface, receive the computed results from Python, and then continue the original execution flow. In this way, the Python-Fortran coupling is concentrated at well-defined interface points, while most of the host model implementation, build structure, and runtime logic remain unchanged.

Nevertheless, this callback-based approach still requires a Python-side registration stage. The callback function is not simply called as an ordinary external routine inside a standalone Fortran executable. Instead, it must first be registered through the F2PY-generated Python module, after which the pre-built Fortran logic can invoke the registered callback during execution. In practice, this means that the PMC still need to initialize the coupling environment, register the Python callback, and launch or coordinate the host-side Fortran entry point. This forms a two-stage workflow: the callback is registered at the Python/F2PY boundary, whereas the actual callback invocation occurs later inside the host-side Fortran execution.

In contrast to the path of PMC organizing PCF modules, the workflow of the path of Fortran program calling FCP functions can be summarized in **Fig. 2b**. While sharing the fundamental logistics of ❶-❹ in **Fig. 2a**, the path of Fortran program calling FCP functions needs to add Python callback statements and interface in ❶-❷. The Fortran interface is defined to specify how the host model communicates with Python-side functions, and the relevant lower-level Fortran routines are modified to call this interface when Python-side computation is required. While ❸ remains nearly unchanged (except for including the callback interface-facing), in ❹, the PMC needs to register callback utility and add `host_entry()` as well as the required import library.

In practice, the two paths described above can of course be used together in the same job if necessary. From the following **Sect. 3** on, we will describe a Latent-3DVar (L3DVar) AI data assimilation scheme that uses autoencoder (AE) to carry out latent space transformation and 3-dimensional variational (3D-Var) minimization to absorb observational information.

## 3 An AE Latent Space 3D-Var (L3DVar) within $H_{f2p}$MDA

In general, in the framework of $H_{f2p}$MDA implemented in **Sect. 2**, any AI algorithm can be fused with Fortran-programmed scientific computation. As application examples of $H_{f2p}$MDA, here we describe a Latent Space 3D-Var (L3DVar) AI data assimilation algorithm that will be applied to a climate model to realize SCDA (**Sect. 4**) and a weather model to fulfill fine-scale Lidar observation assimilation (**Sect. 5**). The L3DVar AI DA algorithm consists of AE latent space transformation and 3D-Var minimization in the latent space (Cheng et al., 2022; Melinc and Zaplotnik, 2024).

### 3.1 The principle of AE latent space transformation

Autoencoders (AEs) are neural representation-learning models that learn a compact latent representation of high-dimensional inputs through an encoder–decoder architecture. In geophysical applications, multiple physical variables can be arranged as separate input channels, enabling the encoder to extract a joint low-dimensional representation of multivariate spatial structures (Hinton and Salakhutdinov 2006; Fan et al. 2025a, b; Fan et al. 2026). Let $\mathbf{x} \in \mathbb{R}^n$ denote a normalized physical-space state vector containing atmosphere, ocean, or other model variables. The trained encoder $E$ maps $\mathbf{x}$ into a latent vector $\mathbf{z} \in \mathbb{R}^r$, and the trained decoder $D$ maps $\mathbf{z}$ back to the physical space:

$$\mathbf{z} = E(\mathbf{x}), \qquad \hat{\mathbf{x}} = D(\mathbf{z}). \tag{1}$$

Here, $\hat{\mathbf{x}}$ is the reconstructed physical-space state. In this study, the AE is trained before data assimilation and then kept fixed during the assimilation procedure. The detailed reconstruction loss, variable normalization strategy, and application-specific training settings are provided in the **Text S2: Implementation details of an AE architecture,** while a schematic workflow of AE-based transformation between the physical space and latent space is shown in **Fig. 3a**.

Variational autoencoders (VAEs), introduced by Kingma and Welling (2013), extend this framework by imposing a probabilistic latent prior, commonly a standard Gaussian distribution, through the Kullback–Leibler divergence term in the variational objective. This regularization can produce a smoother and more structured latent space, but it may also degrade reconstruction accuracy when the regularization is too strong. Because the performance of the subsequent assimilation is directly affected by the fidelity of the decoded analysis fields, we adopt a deterministic AE rather than a VAE in this study. Previous latent-space data assimilation studies have further shown that suitably trained autoencoders can yield latent representations with weakly correlated or approximately diagonal background-error covariance structures, which makes simplified covariance assumptions more plausible for latent-space variational assimilation, including latent-space 3DVar (Fan et al. 2026).

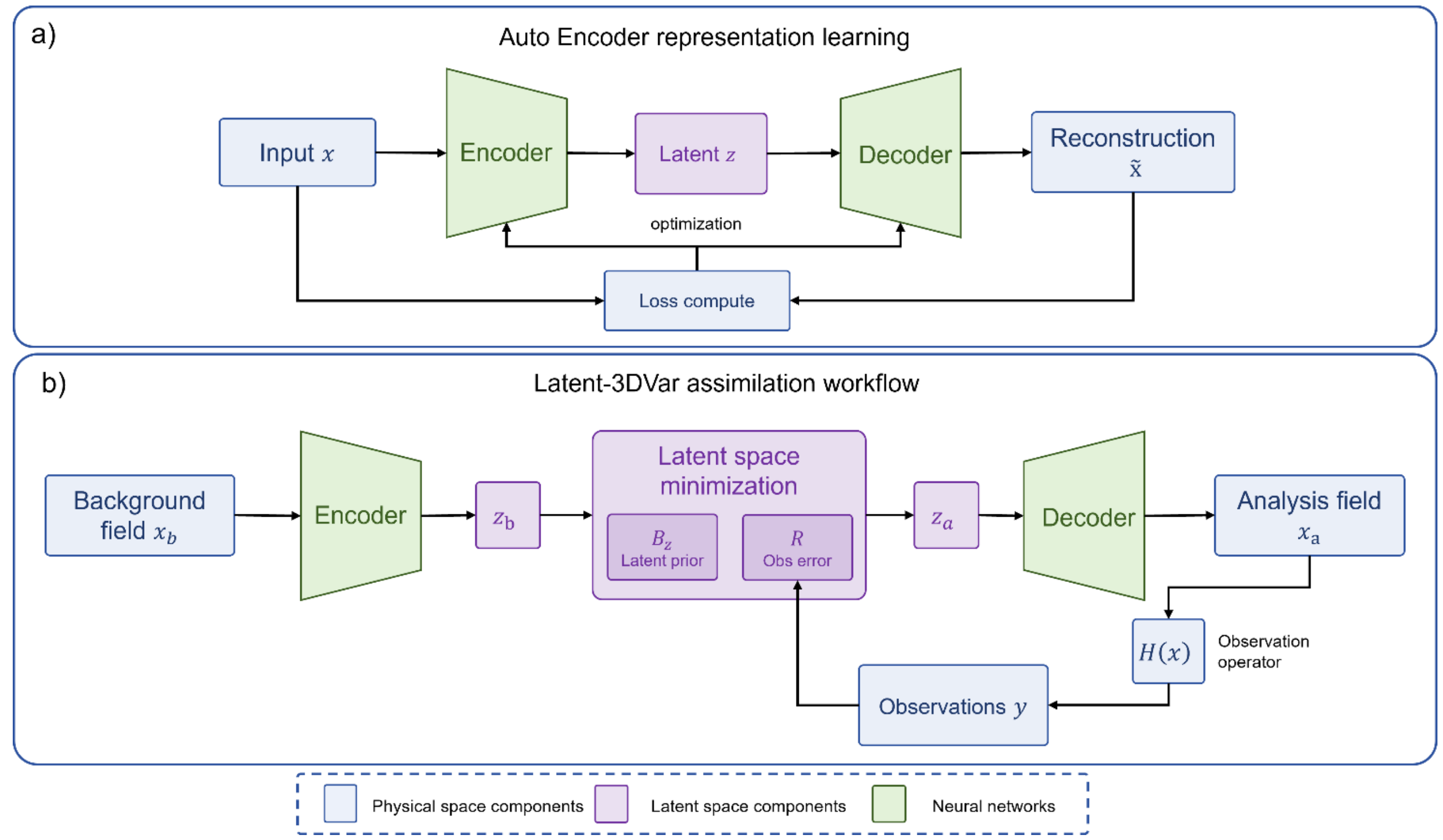

**Figure 3: Schematic workflow of the AE-based Latent Space 3D-Var (L3DVar) algorithm. (a) AE training provides a nonlinear encoder–decoder transformation between the physical model space and a reduced latent space. (b) In the online assimilation stage, the background state $x_b$ is encoded into the latent background vector $z_b$, observational information is incorporated through latent-space 3D-Var minimization, and the optimized latent vector $z_a$ is decoded to produce the physical-space analysis field $x_a$. $B_z$ and R denote the latent-space background-error covariance matrix and the observation-error covariance matrix, respectively.**

### 3.2 Latent space minimization in general

Based on the AE transformation described in **Sect. 3.1**, L3DVar introduces a variational minimization process in the latent space, as illustrated in **Fig. 3b**. Given a background physical-space state $\mathbf{x}_b$, its latent representation is first obtained by the trained encoder as $\mathbf{z}_b = E(\mathbf{x}_b)$. The analysis latent vector $\mathbf{z}_a$ is then estimated by minimizing a 3D-Var-type cost function in the latent space:

$$\mathrm{J}(\mathrm{z}) = \frac{1}{2}\|\mathrm{z} - \mathrm{z_b}\|^2_{\mathrm{B_z^{-1}}} + \frac{1}{2}\|\mathrm{y} - \mathcal{H}(\mathrm{D}(\mathrm{z}))\|^2_{\mathrm{R^{-1}}} \tag{2}$$

In this formulation, $\mathbf{B}_z$ is the background-error covariance matrix in the latent space, $\mathbf{R}$ is the observation-error covariance matrix, $\mathbf{y}$ denotes the observation vector, and $\mathcal{H}$ is the observation operator that maps the decoded physical-space state to the observation space. The two covariance matrices weight and nondimensionalize the background and observation misfit terms, respectively. After minimization, the physical-space analysis state is obtained by decoding the optimized latent vector, i.e., $\mathbf{x}_a = D(\mathbf{z}_a)$.

This formulation differs from conventional physical-space 3D-Var in that the control variable is the reduced latent vector $\mathbf{z}$ rather than the full model state $\mathbf{x}$. Therefore, the dimension of the minimization problem is substantially reduced, while the observation constraint is still evaluated in the physical observation space through the decoded state $D(\mathbf{z})$. Because the decoder is nonlinear and consists of learned neural-network transformations, the resulting observational correction can represent nonlinear cross-variable relationships even though the cost function retains a 3D-Var form. This property is particularly useful for strongly coupled data assimilation, in which observations from one component, such as the atmosphere or ocean, are expected to directly adjust the coupled state of another component through the learned latent representation (Cheng et al., 2024; Lin et al., 2025).

The L3DVar workflow shown in **Fig. 3b** therefore provides a general latent-space assimilation operator. The background state $\mathbf{x}_b$ is encoded into the latent background vector $\mathbf{z}_b$, observational information is incorporated through variational minimization, and the optimized latent state $\mathbf{z}_a$ is decoded back to the physical space as the analysis field $\mathbf{x}_a$. This general workflow is independent of the specific host model and can be coupled to different Fortran-based models through the $\mathrm{H_{f2p}}$MDA infrastructure.

### 3.3 The AE data-training process and 3D-Var latent space minimization for L3DVar

The implementation of L3DVar in $\mathrm{H_{f2p}}$MDA consists of an offline AE training stage and an online assimilation stage. During the offline stage, the encoder $E$ and decoder $D$ are trained using application-specific model states. The basic transformation between physical and latent spaces follows the general AE architecture shown in **Fig. 3a**, whereas the detailed network structure, training samples, normalization procedure, and reconstruction loss are determined by the target application and are therefore documented in the Supplementary Information and cited in the corresponding application sections.

During the online assimilation stage, the trained AE is used as a fixed Python-side AI algorithm within $\mathrm{H_{f2p}}$MDA. At each assimilation time, the background state $\mathbf{x}_b$ provided by the host model is passed to the encoder $E$ to obtain the latent background vector $\mathbf{z}_b$. The latent-space cost function $J(\mathbf{z})$ is then minimized using an efficient PyTorch-based optimizer, such as Adam (Kingma and Ba, 2014). When the prescribed convergence criterion is satisfied, the optimized latent vector $\mathbf{z}_a$ is decoded by $D$ to generate the physical-space analysis field $\mathbf{x}_a$. This analysis field is then returned to the host model and used as the initial condition for the next model integration step.

This implementation design provides the algorithmic connection between the general $\mathrm{H_{f2p}}$MDA infrastructure in **Sect. 2** and the model-specific applications in **Sects. 4 and 5**. In the climate application, the same L3DVar operator is inserted into the coupled atmosphere–ocean data assimilation workflow to realize SCDA. In the weather application, the operator is coupled to a high-resolution model workflow to assimilate fine-scale Lidar observations. Thus, this section defines the application-independent L3DVar algorithm, and the following sections describe how this algorithm is embedded into specific Fortran-based modeling systems through $H_{f2p}$MDA.

## 4 $H_{f2p}$MDA applied to a climate model to realize strongly-coupled data assimilation (SCDA)

### 4.1 The CM2 and its weakly-coupled data assimilation (WCDA) system

The CM2 is a fully-coupled Earth system model consisting of the atmosphere, ocean, land and sea-ice components (Lin, 2004; Gnanadesikan et al., 2006; Winton, 2000), having been an important member of the Coupled Model Intercomparison Project (CMIP) (Delworth et al., 2006b; Randall, 2007; Taylor et al., 2012; Eyring et al., 2016). The version we used in this study has the atmosphere model with 2° latitude × 2.5° longitude horizontal resolution and 24 vertical levels, and the ocean with 1°×1° horizontal resolution and 50 vertical levels, time stepping with 30 minutes and 2 hours for atmosphere and ocean as well as their coupling. As the first CGCM's CDA system, the first version of CM2-CDA was developed based on the ensemble Kalman filter (EnKF) (Zhang and Anderson, 2003; Zhang et al., 2007), implemented in a weakly-CDA (WCDA) manner, i.e., the atmosphere and ocean conduct their DA in their own model components. However, an EnKF CDA system is very expensive and has a limitation on extracting low-frequency observational signals (Yu et al., 2019). To resolve this issue, based on scale disassembling of a long time series of single model solution and the EnKF's framework, a multiscale high-efficiency approximate EnKF (MSHea-EnKF) was designed (Yu et al., 2019) and implemented into the CM2 (Lu et al., 2023). Therefore, the current version of CM2-CDA employed in this study uses a sequential WCDA procedure which forwards on single CM2 coupled model integrated with atmosphere-ocean observations as they are available. In the way of MSHea-EnKF, instead of forwarding the coupled model in an ensemble manner, the multiscale background ensembles are initially constructed from a long time series of single model solution but innovated with the ongoing CDA procedure.

In a WCDA procedure, the atmosphere (ocean) observations only adjust the atmosphere (ocean) its own state and observational information is transferred cross the atmosphere and ocean by coupled model exchanged fluxes. Without direct observational adjustment between the atmosphere and ocean, imperfect coupling physics may still introduce errors into the coupled model and CDA remains in an incomplete fashion. The strongly-CDA (SCDA) has been pursued to fulfill sufficient balanced and coherent coupled state estimation in CDA (Zhang et al., 2020) but it remains challenging (e.g. De Rosnay et al., 2022) imposed by the limitation of linear regression in traditional DA algorithms for such different characteristic scales cross the atmosphere and ocean. The natural nonlinear nature of machine deep learning AI algorithms makes SCDA feasible if it is inserted into a WCDA system as an additional term.

Next, we will describe how to apply the $H_{f2p}$MDA framework outlined in **Sect. 2** so that an AI SCDA scheme can be fused with the legacy CM2-CDA to form an $H_{f2p}$MDA-$CM2_{SCDA}$.

### 4.2 $H_{f2p}$MDA fusing L3DVar with CM2-CDA to form $H_{f2p}$MDA-$CM2_{SCDA}$

In general, the CM2 model has a very clear modular feature, in which the atmosphere and ocean components have relatively-independent structure, with land as atmospheric boundary processes and sea-ice being the interface of atmosphere and ocean. We use the path of PMC organizing PCF modules (as shown in **Fig. 2*a***) to implement $H_{f2p}$MDA-$CM2_{SCDA}$. The main controller *coupler_main* organizes all interfaces between model components and loops time integration as every half hour for the atmosphere and every two hours for the ocean and atmosphere-ocean coupling. The weakly CDA conducts atmosphere (ocean) DA computation within the atmosphere (ocean) component and it therefore does not change the basic logic structure of *coupler_main.* The entire CM2-CDA package of codes consists of 4 static libraries: *libfms.a*, *libland.a*, *libecda.a* and *libcoupler.a*. Here *libfms.a* contains all the lowest shared information such as MPI and remapping functions etc. and *libland.a* is a relatively-independent application package for land model based on *libfms.a*, while *libecda.a* includes all atmosphere and ocean model components as well as CDA subroutines beside land model. Finally, *libcoupler.a* is the highest infrastructure to connect all parts above together by *coupler_main*. To make all Fortran modules in these static libraries becoming PCF modules, we first partition the main controller *coupler_main.F90* as two parts, *cm2_cda_mainsubs.F90* and *cm2_cda_plugs.F90.* The *cm2_cda_mainsubs.F90* is basically a modified version of *coupler_main.F90*, in which the main program is disassembled into a list of subroutines such as *cm2_cda_maininit()*, *atmos_step()* and *ocean_step()* etc. as module initialization and single step model integration. The *cm2_cda_plugs.F90* simply organizes various model component integration of single step as plug-ins (*tool_atmos_step()* and *tool_ocean_step()* etc., for instance) for PMC.

Following the procedure described in **Sect. 2**, we first recompile 4 static libraries *libfms.a*, *libland.a*, *libecda.a* and *libcoupler.a* by adding "-fPIC" option tag. Then we use *f2py* to combine *cm2_cda_plugs.F90* with these 4 static libraries together to create the signature file *cm2_cda.pyf* and DLL *cm2_cda.python-311-x86_64-linux-gnu.so*. Finally, the DLL *cm2_cda.python-311-x86_64-linux-gnu.so* is copied to work directory where *cm2_cda_main.py* is stored.

Based on the guideline of PMC's structure described in **Text S1 Step-4** and **Sect. 2.2.1**, we construct the PMC of $H_{f2p}$MDA-$CM2_{SCDA}$, called *cm2_cda_main.py.* In this case, as arranging time integration of the coupled model, *cm2_cda_main.py* also is responsible for connecting the atmosphere-ocean interface with the AI SCDA algorithm implemented by L3DVar described in **Sect. 3**. Then following the procedure described in **Text S1 Step-4**, we can run *cm2_cda_main.py* to complete the process of $H_{f2p}$MDA-$CM2_{SCDA}$'s design and execution.

Since the PMC *cm2_cda_main.py* conducts the coupled model time integration and it explicitly controls the evolution of coupling interface between the atmosphere and ocean, we can conveniently insert SCDA into the model integration. Next, we will evaluate the results of AE latent space reconstructing coupling variables as well as resulted SCDA by minimization in the latent space.

### 4.3 Test results of SCDA by $H_{f2p}$MDA-$CM2_{SCDA}$

#### 4.3.1 The AE reconstruction of physical variables at air-sea interface by latent space

In the $H_{f2p}$MDA-$CM2_{SCDA}$ system, to produce SCDA results, we first train the encoder and decoder to reconstruct the reduced-order state in latent space using the atmosphere and ocean state variables at the air-sea interface. High-efficiently reconstructing the latent space with necessary accuracy is the core of training scheme which consists of various selections on parameters and loop levels as well as neural networks etc. For the CM2-SCDA, we set the input vector $x_0$ consisting of the atmosphere surface wind, temperature and pressure ($U_s$, $V_s$, $T_s$, $P_s$), and ocean surface currents, temperature and height (SSU, SSV, SST, SSH). The concrete architecture design and training process are described in **Text S2.1 and S2.2**. However, it is worth to mention that different from a general case in which the physical state usually only includes information within a component of multi-sphere geofluid, in this SCDA case, the physical state being transformed consists of atmospheric and oceanic information with different characteristic scales. Under this circumstance, the AE training process therefore includes a Cross Attention step which consists of mixing and its inverse (separation). Apart from this point, the AE still follows a pretty general training procedure for downsampling and upsampling.

All variables are normalized by their corresponding variable ranges before training. The training dataset is constructed from a 50-year free-running CM2 simulation sampled every 3 h (totally 14600 samples in time). The validation set contains 16060 samples and is used to evaluate the reconstruction accuracy. Because the main objective of this study is to evaluate the feasibility of latent-space data assimilation rather than to develop a general-purpose reconstruction model, the independent assimilation cases are treated as the practical out-of-sample tests.

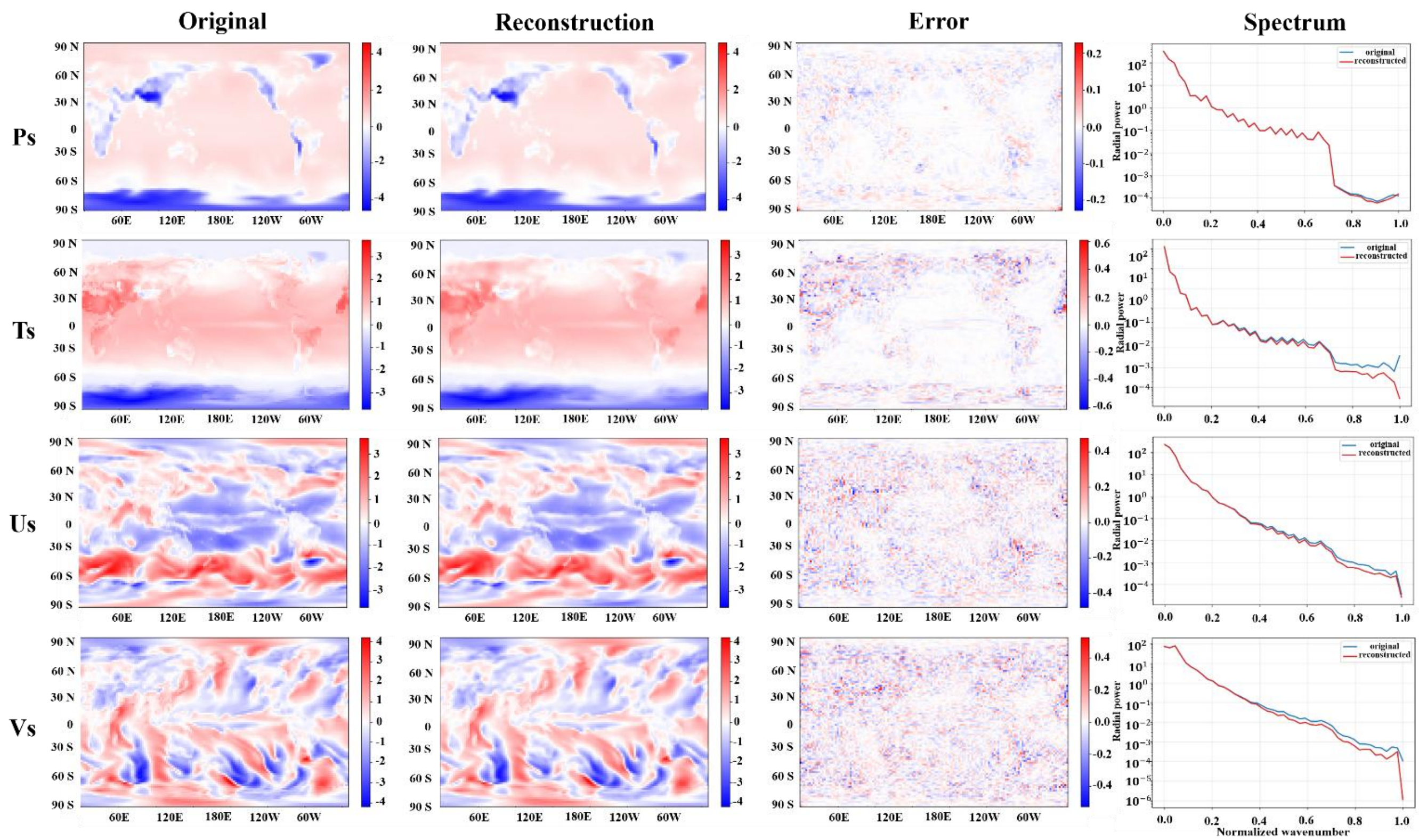

**Figure 4: Reconstruction of CM2 atmospheric air–sea interface variables in the AE latent space. Rows show surface pressure (Ps), surface temperature (Ts), surface zonal wind (Us) and surface meridional wind (Vs). Columns show the original normalized field, the reconstructed field decoded from the latent vector, the reconstruction error and the radially averaged power spectrum of the original and reconstructed fields.**

**Figures 4 and 5** show representative reconstruction examples for the atmospheric and oceanic variables, respectively. The reconstructed fields (middle-left panels) preserve the main large-scale structures of the original variables (leftmost panels), and the corresponding errors (middle-right panels) remain small compared with the normalized field amplitudes. The spectra (rightmost panels) further show that the reconstructed fields retain the dominant low-wavenumber structures and follow the original spectral slopes over most resolved scales, although slight smoothing is still visible at the highest wavenumbers. This behavior is expected for an encoder–decoder representation and does not prevent the reconstructed fields from providing a physically reasonable background for latent-space assimilation.

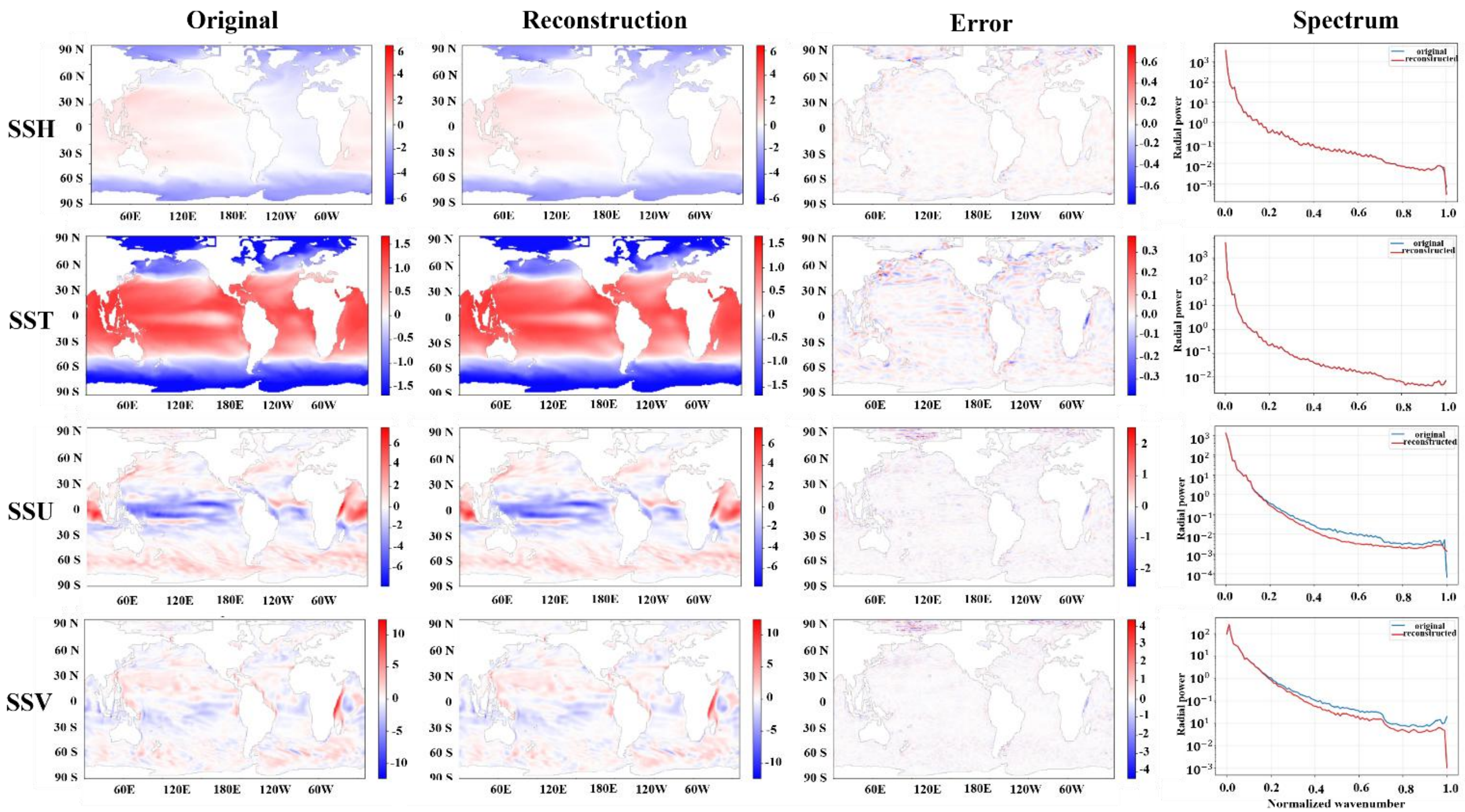

**Figure 5: Same as Fig. 4, but for CM2 oceanic air–sea interface variables, including sea surface height (SSH), sea surface temperature (SST), surface zonal current (SSU) and surface meridional current (SSV).**

The quantitative metrics in **Table 1** confirm this visual assessment. Here, MSE, MAE and RMSE denote the mean squared error, mean absolute error and root mean squared error of the normalized fields, respectively. SSIM denotes the structural similarity index, and PSNR denotes the peak signal-to-noise ratio. The total SSIM reaches 0.9945, indicating that the AE preserves the spatial structure of the coupled air–sea interface variables. Among the selected variables, the ocean current components have relatively larger errors than the thermodynamic variables, reflecting their stronger small-scale variability. Overall, the reconstruction accuracy is sufficient for the following SCDA experiment.

**Table 1: Validation-set reconstruction metrics for CM2 air–sea interface variables in the AE latent space. The metrics are calculated from 16060 validation samples after variable normalization. "Total" denotes the metric averaged over all selected atmospheric and oceanic variables.**

| Variables | MSE | MAE | RMSE | SSIM | PSNR |
|---|---|---|---|---|---|
| Ps | 0.0002 | 0.010 | 0.014 | 0.9996 | 51.53 |
| Ts | 0.0035 | 0.035 | 0.059 | 0.9977 | 39.10 |
| Us | 0.0040 | 0.045 | 0.063 | 0.9977 | 41.29 |
| Vs | 0.0054 | 0.053 | 0.074 | 0.9967 | 41.35 |
| SSH | 0.0013 | 0.024 | 0.036 | 0.9991 | 46.27 |
| SST | 0.0008 | 0.018 | 0.027 | 0.9992 | 40.88 |
| SSU | 0.0148 | 0.079 | 0.120 | 0.9918 | 38.90 |
| SSV | 0.0320 | 0.117 | 0.178 | 0.9854 | 39.90 |
| Total | 0.0109 | 0.056 | 0.104 | 0.9945 | 40.11 |

### 4.3.2 Latent-space evaluation and $\mathbf{B}_z$-related correlation diagnosis

After verifying the reconstruction accuracy, we further examine whether the latent variables provide a suitable control space for variational assimilation. In L3DVar, the background-error covariance matrix $\mathbf{B}_z$ weights the departure of the analysis latent vector from the background latent vector. A simplified $\mathbf{B}_z$ is only useful when the latent variables are not strongly correlated. Here we adopt an NMC-like method (Parrish and Derber, 1992) to calculate the correlation of temporal differences to approximate the $\mathbf{B}_z$.

**Figure 6** shows three complementary diagnostics. The channel-correlation matrix indicates that the latent channels are dominated by diagonal elements, although weak off-diagonal structures remain. These residual correlations are not unexpected, because the CM2 latent vector represents coupled atmospheric and oceanic interface variables with different spatial scales. The spatial-correlation matrix for a representative latent channel also shows a clear diagonal dominance, suggesting that the latent representation does not introduce strong long-range artificial dependence. Finally, the distribution of off-diagonal correlations is sharply centered near zero, with 95% of absolute off-diagonal correlations smaller than about 0.290.

These diagnostics indicate that the trained latent space is sufficiently decorrelated for the present demonstration. The near-diagonal structure provides practical support for using a simplified $\mathbf{B}_z$ like a diagonal matrix in the latent-space minimization.

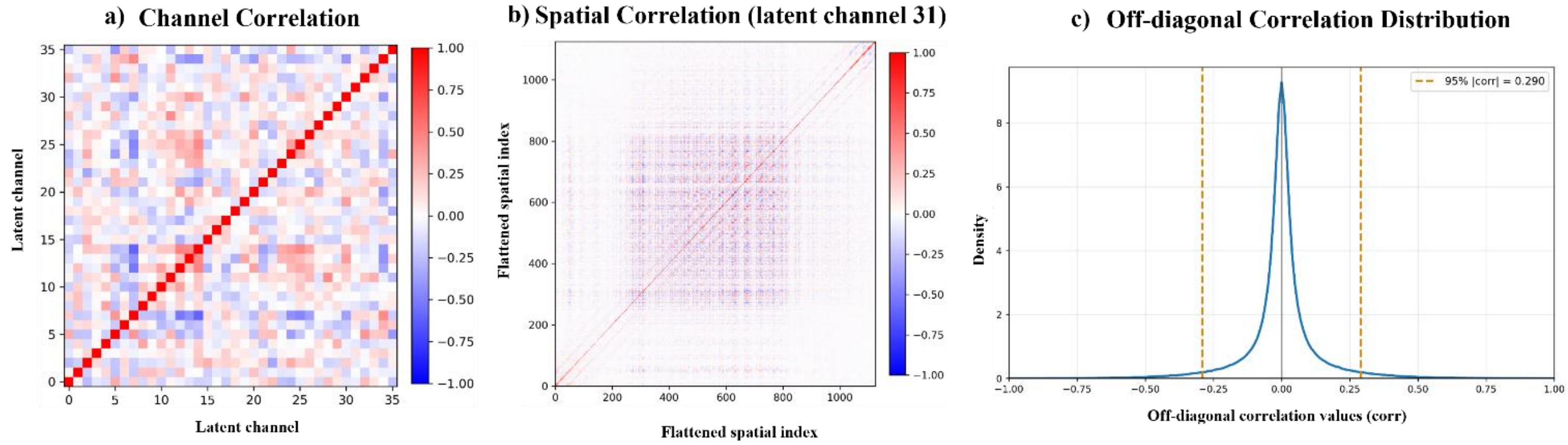

**Figure 6: Latent-space correlation diagnosis for the CM2 AE representation. (a) Correlation matrix among latent channels. (b) Spatial correlation matrix for a representative latent channel. (c) Probability density distribution of off-diagonal correlation coefficients. The dashed vertical lines mark the 95% range of absolute off-diagonal correlations.**

### 4.3.3 The SCDA results

In the case of $H_{f2p}$MDA-CM2$_{SCDA}$, through the AE training, we establish a latent space which is transformed from the air-sea interface space so that we can efficiently conduct minimization to blend atmospheric and oceanic observations into the model space. Here while the size of coupled state vector in physical space is 339840, through space transformation, the state vector size in the latent space becomes 40500 with a reduction rate of 88%, which makes fast and efficient minimization feasible. Next, we will show that although only 12% of physical space vector size is used in latent space minimization, the reconstruction fields are reasonably representing the background physical information for extracting observational information.

In this test case, we use the ERA5 (5$^{th}$ generation atmospheric reanalysis of European Centre for Medium-Range Weather Forecasts) surface pressure Ps (with a 25 km horizontal resolution) and OISST (NOAA 1/4°Daily Optimum Interpolation Sea Surface Temperature) as atmospheric and oceanic observations respectively. With the usual observational errors of 5 hPa (for Ps) and 0.5 $^{o}$C (for SST), we run the $H_{f2p}$MDA-CM2$_{SCDA}$ system on the whole year of 1982 and present a typical minimization process and the whole year verification results in **Figs. 7*a-b***. The latent space can be comprehended as a reduced-order holographic space that represents all nonlinear complex relationships of physical variables. In this space, as minimization proceeds (see details in **Text S3**), when the observation loss decreases, the latent space state vector $z$ gets adjusted as the background loss increases so that the observations are coherently incorporated into the background at the convergent point that the background and observation reach equilibrium.

**Figures. 7*a-b*** show that due to the low-frequency and quasi-linear nature of ocean motions, both WCDA and SCDA make the model SST quickly converge to observations and keep in nearly RMSEs in most of the time. In the contrary, because of strong internal variability of the atmosphere, the curves of Ps RMSEs in WCDA and SCDA quickly separate after a few days, but SCDA does not show its advantage within a few months. After about 4~5 months, SCDA starts to show a little smaller RMSEs. In terms of a 6-month mean in July to December, SCDA reduces the WCDA's RMSE by roughly 4%, in which SCDA's RMSE

is 4.62 out of 4.81 of WCDA's. The examples of spatial distributions of differences of SST and $P_s$ RMSEs between SCDA and WCDA are given in **Figs. 7*c-d***, which show that the large effects of SCDA occur mainly in the tropics for SST and middle-high latitude storm track area for $P_s$. More examples for SST produced by SCDA and WCDA and their differences are shown in **Fig. 8**.

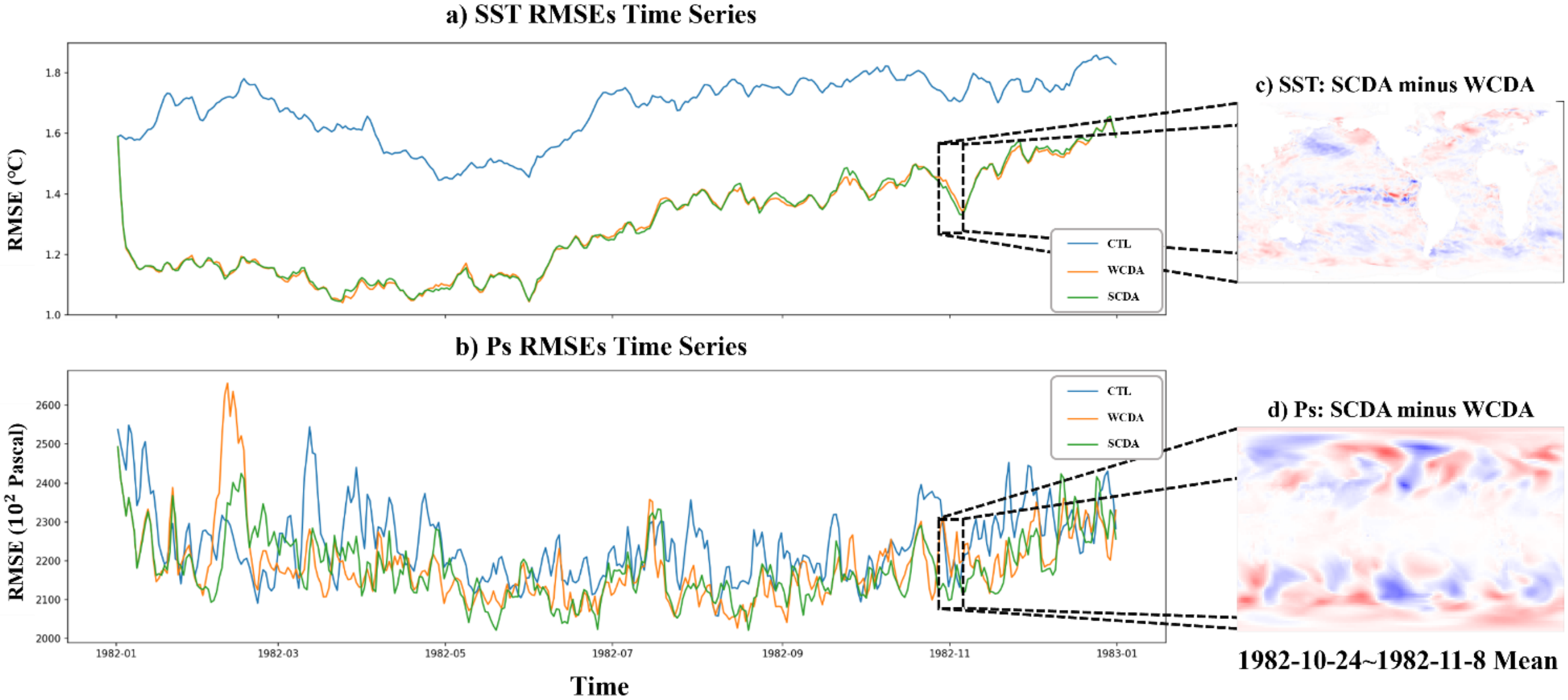

**Figure 7: ab) The time series of root mean squared errors (RMSEs) of SST and Ps in free model control simulation (CTL) (blue), and WCDA (orange) as well as SCDA (green) produced by Hf2pMDA-CM2$_{CDA}$ at the 200$^{th}$ iteration of minimization. The WCDA is a multiscale high-efficient approximate filter (MSHea-EnKF). cd) Distributions of the mean differences of RMSEs of SST and Ps in SCDA and WCDA (SCDA-WCDA) in the sub-time series from 1982-10-24 0130Z to 1982-11-8 0130Z (denoted by the dashed-black box in panels *a-b*).**

**Figure 8** indicates that SCDA can produce large effects on the tropical instability waves (TIWs) since very high-frequency atmosphere-ocean flux exchanges occur and easily render strong nonlinearity. It's worth to mention that in this $H_{f2p}$MDA-CM2$_{CDA}$ case, the model resolution is pretty low (~200 km for the atmosphere and ~100 km for the ocean). Under the circumstance, the coupled model dynamics in general is quasi-linear (Perlwitz et al., 2017), and it is possible that WCDA can extract the most of observational information into the coupled model. However, our preliminary results using $H_{f2p}$MDA-CM2$_{CDA}$ exhibit the promise that after a sufficient time scale, SCDA can create additional values to the CDA system, especially for the active air-sea interaction (TIWs, for instance) regions. This may become critically-important as the model attempts to resolve fine-scale and frequent air-sea coupling processes as a tropical cyclone or mesoscale eddy passes by.

The results above show that in this test case, although only 6% of physical space vector size is used in latent space minimization, the reconstruction fields are able to represent the background physical information for extracting observational information.

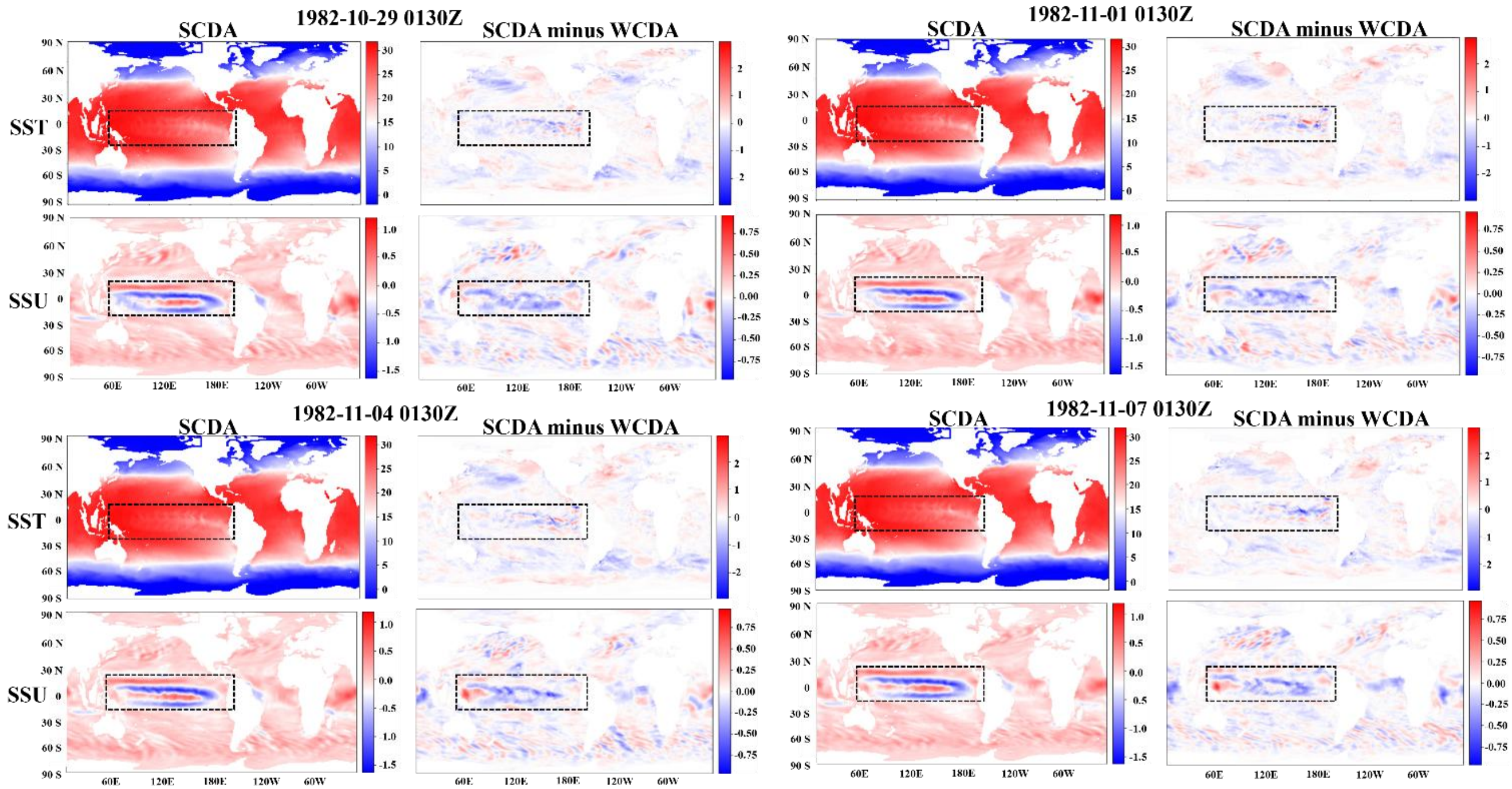

**Figure 8: The physical samples of SST and SSU at the four times. The dashed-black box in each panel marks the area of active tropical instability waves where the SCDA easily renders difference from WCDA as shown by the middle-left and most-right columns denoted as SCDA minus WCDA. 4 sets of SCDA SST and SSU and their increments from WCDA at 1982-10-29 0130Z, 1982-11-01 0130Z, 1982-11-04 0130Z and 1982-11-07 0130Z are shown in the upper-left, upper-right, lower-left and lower-right.**

## 5 $H_{f2p}$MDA applied to a weather model to realize fine-scale convection data assimilation (FSDA)

### 5.1 WRF and its multiscale data assimilation system

Here we use a version of Weather Research and Forecasting (WRF, v3.7.1) model configured as 3-layer nesting at 9, 3 and 1 km horizontal resolutions for the West of China (D01), Central Sichuan Province (D02) and Chengdu City (D03) respectively as shown in **Fig. 9*a***. The WRF model is divided into 50 vertical levels from the surface to the model top at 50 hPa. The model includes most of the primary physics such as the Kain-Fritsch convection parameterization (Kain, 2004), Rapid Radiative Transfer Model for GCMs (RRTMG) longwave and shortwave radiations (Clough et al., 2005) and Yonsei University (YSU) boundary layer (Hu et al., 2013) as well as WRF single-moment 3-class (WSM3) microphysics (Hong et al., 2004) etc. The initial and boundary conditions are drawn from the ERA5 reanalysis dataset in this study. The original purpose of this WRF configuration is to study the impact of lidar observations on the Chengdu City. Here we employ it to exhibit the capability to conveniently carry out parallel experiments using $H_{f2p}$MDA conducting traditional DA and AI L3DVar.

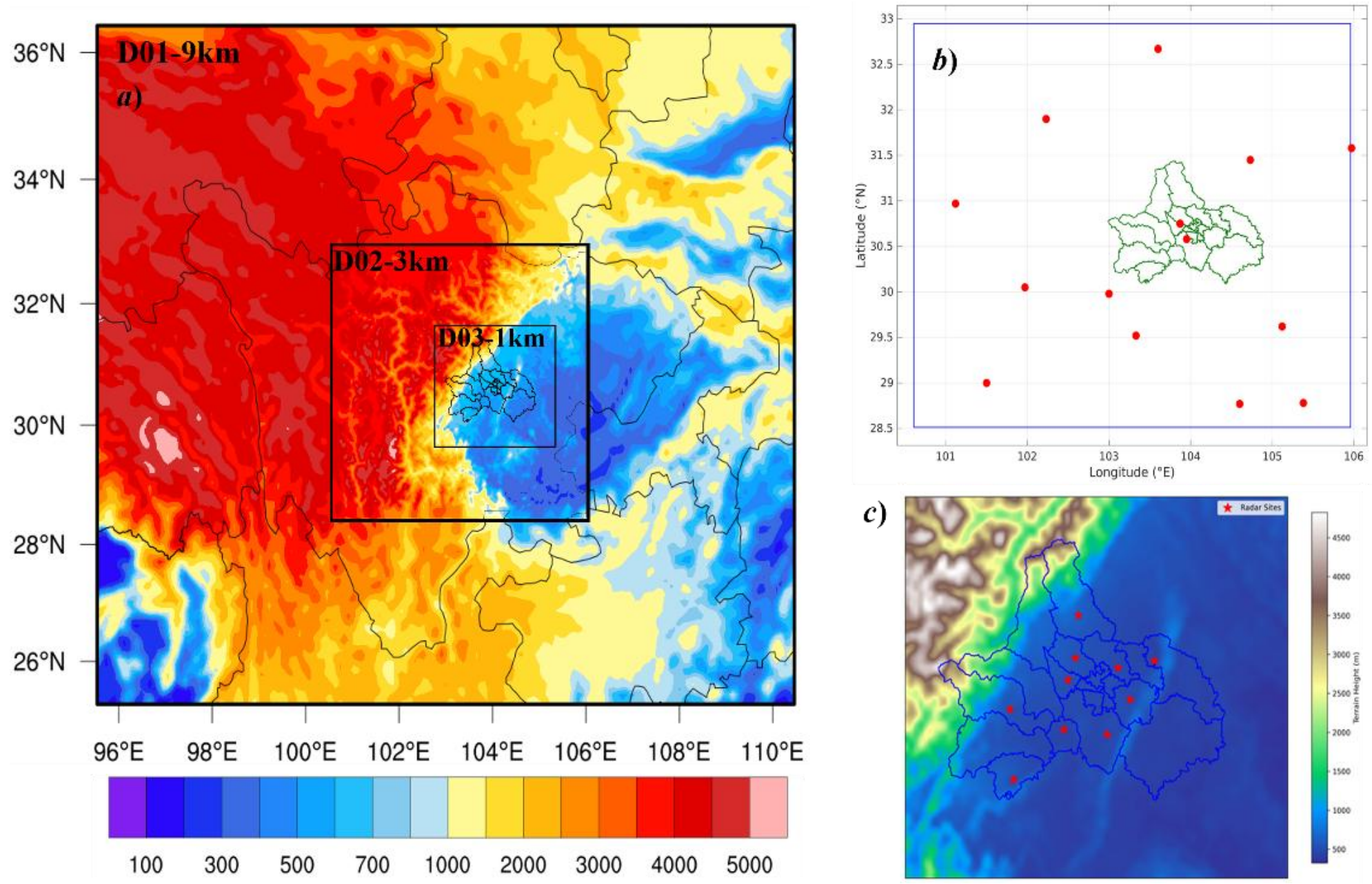

**Figure 9: The configurations of *a*) 3-layer nesting model domain at 9, 3 and 1 km horizontal resolutions for the West China (D01), Central Sichuan Province (D02) and Chengdu City (D03), *b*) conventional observation stations in D02, and *c*) Lidar stations deployed in Chengdu City D03.**

Same as the CM2-CDA described in **Sect. 3.1**, a multiscale high-efficiency approximate EnKF (MSHea-EnKF) (Yu et al., 2019) data assimilation algorithm is used in this multi-layer nesting WRF to carry out multiscale observational information extraction (Wang et al., 2024). To do that, we conduct the large scale "observational" constraint in D01 using ERA5's wind and temperature, local data constraint in D02 and D03 using the conventional observations shown in **Fig. 9*b*** (in D02) as well as additional Lidar observations shown in **Fig. 9*c*** (in D03). In this study, once the $H_{f2p}$MDA-WRF$_{FSDA}$ system (will be described in the next section) is set, we use it to conduct parallel assimilation experiments in the 1 km resolution Chengdu City core domain by the classic MSHea-EnKF and AI L3DVar to test validation of the $H_{f2p}$MDA-WRF$_{FSDA}$.

## 5.2 $H_{f2p}$MDA fusing L3DVar with WRF$_{DA}$ to form $H_{f2p}$MDA-WRF$_{FSDA}$

Unlike the CM2 model which is highly modularized, the WRF model uses a recursive function called *integrate* to perform downscaling simulations by inputting differently-defined domain as its argument. Under this circumstance, we use the path of Fortran program calling FCP functions (as shown in **Fig. 3*b***) to implement $H_{f2p}$MDA-WRF$_{FSDA}$. We adopt a strategy in which the WRF main program is changed to a module containing subroutine *wrf* and the whole model and its DA codes are packed as an entity that is used by *wrfda_plug.F90*. In this case, the *wrfda_plug.F90* only defines a subroutine called *call_wrf_main* which calls the subroutine *wrf* and defines an *external python_foo* before the subroutine call. The *python_foo* is passed in by the *call_wrf_main* as an argument from the PMC that will be described below. The external *python_foo* is a trained LDA algorithm which is used by *callback_python.F90* where a *python_da* interface is defined as a Fortran subroutine which is public to any Fortran module if applicable.

Once the preparation described above is done, we follow the procedures described in **Text S1** to complete environment setting and Fortran codes compiling. At this point, the signature file *wrfda.pyf* and the "shared object" *wrfda.cpython-311-x86_64-linux-gnu.so* are ready for being imported to PMC.

In this case, the PMC *wrfda_main.py* is very concise. After importing necessary Fortran and Python modules including some interfaces and utilities for usage of L3DVar, the *wrfda_main.py* directly makes a Python call to *call_wrf_main* as *wrfda*.*wrfda_plug*.*call_wrf_main*(*transpond*). Here the *transpond* is the function that will become external *python_foo* in the subroutine *call_wrf_main* by argument passing. Following the procedure described in **Sect. 2.3.2**, we can execute the $H_{f2p}$MDA-$WRF_{DA}$ system.

### 5.3 Test results of FSDA by $H_{f2p}$MDA-$WRF_{FSDA}$

#### 5.3.1 The AE reconstruction of WRF D03 physical variables by latent space

We use ERA5 dataset to create boundary and initial conditions and conduct 3-layer nesting downscaling simulations as described in **Sect. 5.1**. Each case is run for 36 hours with hourly output in D03. We discard the first 12 hours as downscaling spinup and connect all data in last 24 hours of these cases in 2022-2024 to form a 3-year hourly dataset (totally 26280 samples in time) for the AE reconstruction training. The architecture design and training process are described in details by **Text S2.3 and S2.4**. It's worth to mention that in this case, pursuing 1 km resolution high-precision expression, we particularly concern local small scale information, and therefore we set a relatively large latent_dim (196608) with only a reduction rate of roughly 25%.

In this test case, we only assimilate the surface observations as shown in **Figs. 9*b-c***. We first show the reconstruction accuracy. As examples, the reconstructed  moisture ($q_1$), temperature ($T_1$) and surface wind ($u_1$, $v_1$) fields are shown in **Figs. 10**, which are randomly drawn from a 1-month validation dataset of June 20 - July 20, 2025, for which we will conduct parallel MSHea-EnKF and AI L3DVar experiments using $H_{f2p}$MDA-$WRF_{DA}$ in D03. **Figure 10** shows that the reconstructed fields reproduce the main spatial structures of the original normalized fields. The error maps (next-right column) show that reconstruction errors are generally small and spatially localized. The spectra (most-right column) further indicate that the reconstructed fields preserve the main spectral distribution of the original fields, although weak high-wavenumber smoothing remains.

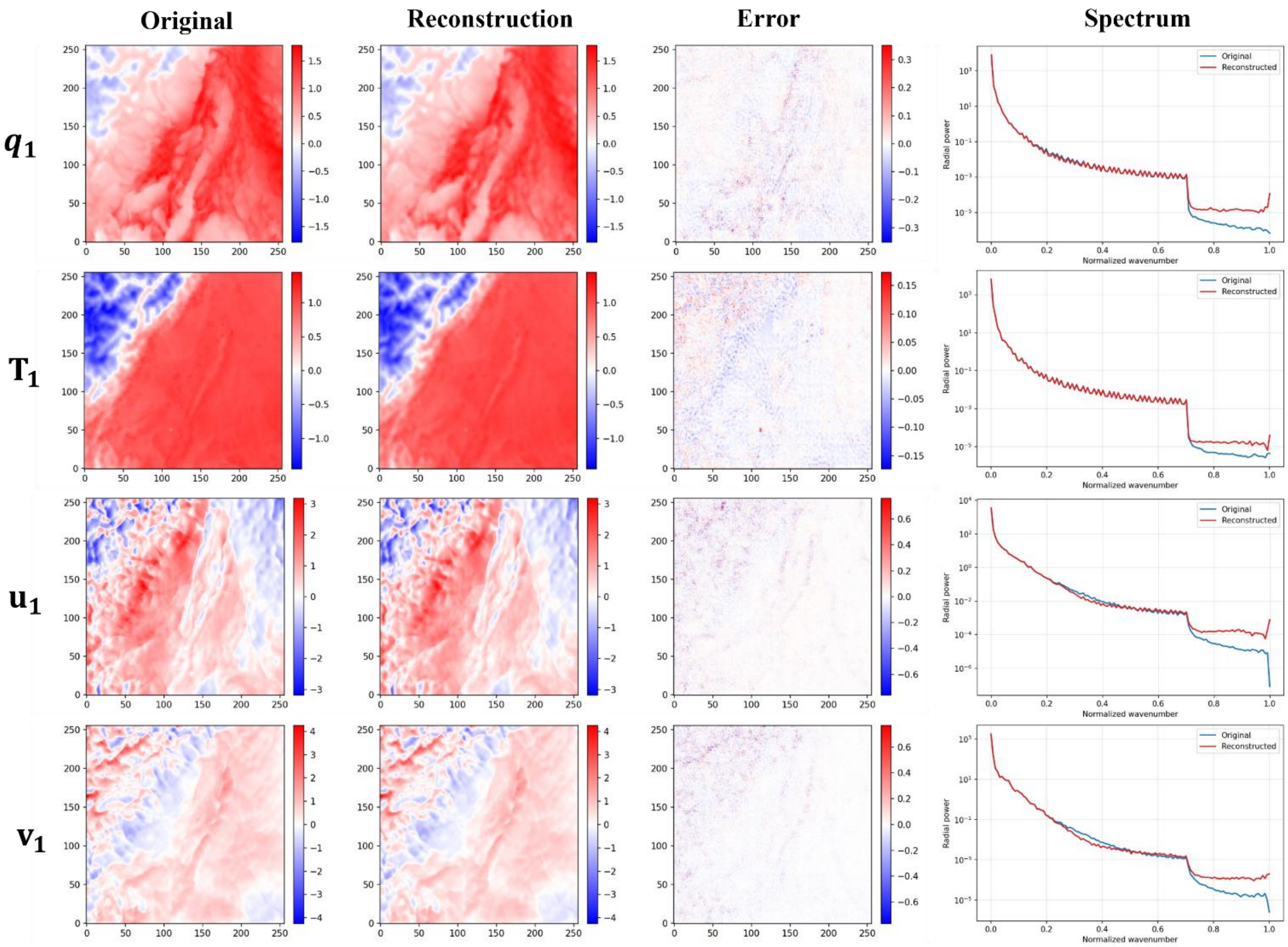

**Figure 10: Reconstruction of WRF D03 first model-level wind variables in the AE latent space. Rows show $q_1$, $T_1$, $u_1$ and $v_1$. Columns show the original normalized field, the reconstructed field decoded from the latent vector, the reconstruction error and the radially averaged power spectrum of the original and reconstructed fields. The sample is taken from 00:00 UTC 23 June 2025.**

The validation metrics are summarized in **Table 2**. The total SSIM is 0.9966, indicating that the AE reconstruction retains the spatial structure of the D03 near-surface variables. The small RMSE and MAE values further show that the reconstruction error is limited after normalization. These results suggest that the trained latent representation is sufficiently accurate for the following D03 latent-space assimilation experiment.

**Table 2. Validation-set reconstruction metrics for WRF D03 near-surface variables in the AE latent space. The metrics are calculated from 2183 validation samples after variable normalization. "Total" denotes the metric averaged over $q_1$, $T_1$, $u_1$ and $v_1$.**

| Variables | MSE | MAE | RMSE | SSIM | PSNR |
|---|---|---|---|---|---|
| $q_1$ | 0.00018 | 0.0089 | 0.0127 | 0.9968 | 40.68 |
| $T_1$ | 0.00028 | 0.0107 | 0.0164 | 0.9989 | 42.91 |
| $u_1$ | 0.00387 | 0.0335 | 0.0597 | 0.9952 | 42.52 |
| $v_1$ | 0.00298 | 0.0314 | 0.0528 | 0.9954 | 42.76 |
| Total | 0.00183 | 0.0211 | 0.0427 | 0.9966 | 42.60 |

### 5.3.2 Latent-space evaluation and $B_z$-related correlation diagnosis

The latent-space correlation structure is also examined for the WRF D03 application. This diagnosis is used to evaluate whether the latent variables provide a weakly correlated control space for L3DVar. As in the CM2 case, the purpose is to assess whether a simplified $\mathbf{B}_z$ is acceptable for the present demonstration.

**Figure 11** shows the latent-space correlation diagnostics for the trained WRF AE. Both the channel-correlation and representative spatial-correlation matrices are strongly dominated by their diagonal elements. Compared with the CM2 case, the off-diagonal structures are weaker, which indicates a stronger decorrelation in the WRF latent representation. The off-diagonal correlation distribution is sharply peaked around zero, and 95% of absolute off-diagonal correlations are smaller than about 0.117. This result supports the use of a simplified latent-space $\mathbf{B}_z$ in the D03 LDA experiment.

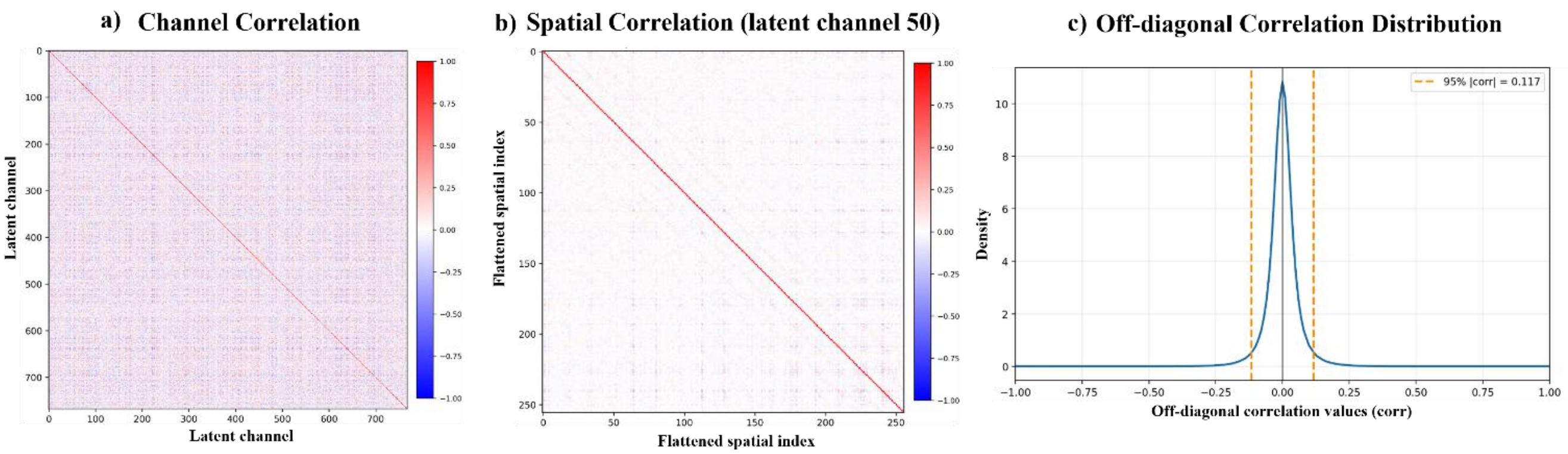

**Figure 11. Latent-space correlation diagnosis for the WRF D03 AE representation. (a) Correlation matrix among latent channels. (b) Spatial correlation matrix for a representative latent channel. (c) Probability density distribution of off-diagonal correlation coefficients. The dashed vertical lines mark the 95% range of absolute off-diagonal correlations.**

### 5.3.3 The FSDA results in D03

Once the latent space AE is trained, the minimization in the latent space can be conducted with the Lidar observations in the D03 to carry out L3DVar. We insert the L3DVar into the $WRF_{DA}$ system under $H_{f2p}$MDA to replace the D03 DA procedure form $H_{f2p}$MDA-$WRF_{FSDA}$ system. Setting an optional switch, we can perform DA experiments with classic multiscale $WRF_{DA}$ and new L3DVar-$WRF_{FSDA}$ within the $H_{f2p}$MDA-$WRF_{FSDA}$ framework in a parallel fashion. We insert the AE encoder and decoder trained in **Text S2.3 and S2.4** and employ the latent space minimization described in **Text S3** to replace the D03 DA procedure. In this way, we can perform DA experiments with classic multiscale MSHea-EnKF and new AI L3DVar within the $H_{f2p}$MDA-$WRF_{DA}$ framework in a parallel fashion. With the observational error as 1 m/s (for wind), 1 ºC (for atmospheric temperature), we run the system from 00UTC 21 June 2025 to 00UTC July 21 2025 with initial and boundary conditions remapped from ERA5 reanalysis, in which the WRF model with 9/3 (1) km resolutions in D01/D02 (D03) assimilates conventional surface station (Lidar) observations shown in **Figs. 9*b-c***. We first show the distribution of analysis increments at the first step assimilation in Fig. 12. We see very local $u_1$ and $v_1$ adjustments that distribute around Lidar locations (**Figs. 12*a-b***), and although the Lidar only provides wind measurements in this case, the moisture and temperature also get consistent adjustments (**Figs. 12*c-d***) under this AE-LDA multivariate nonlinear assimilation framework.

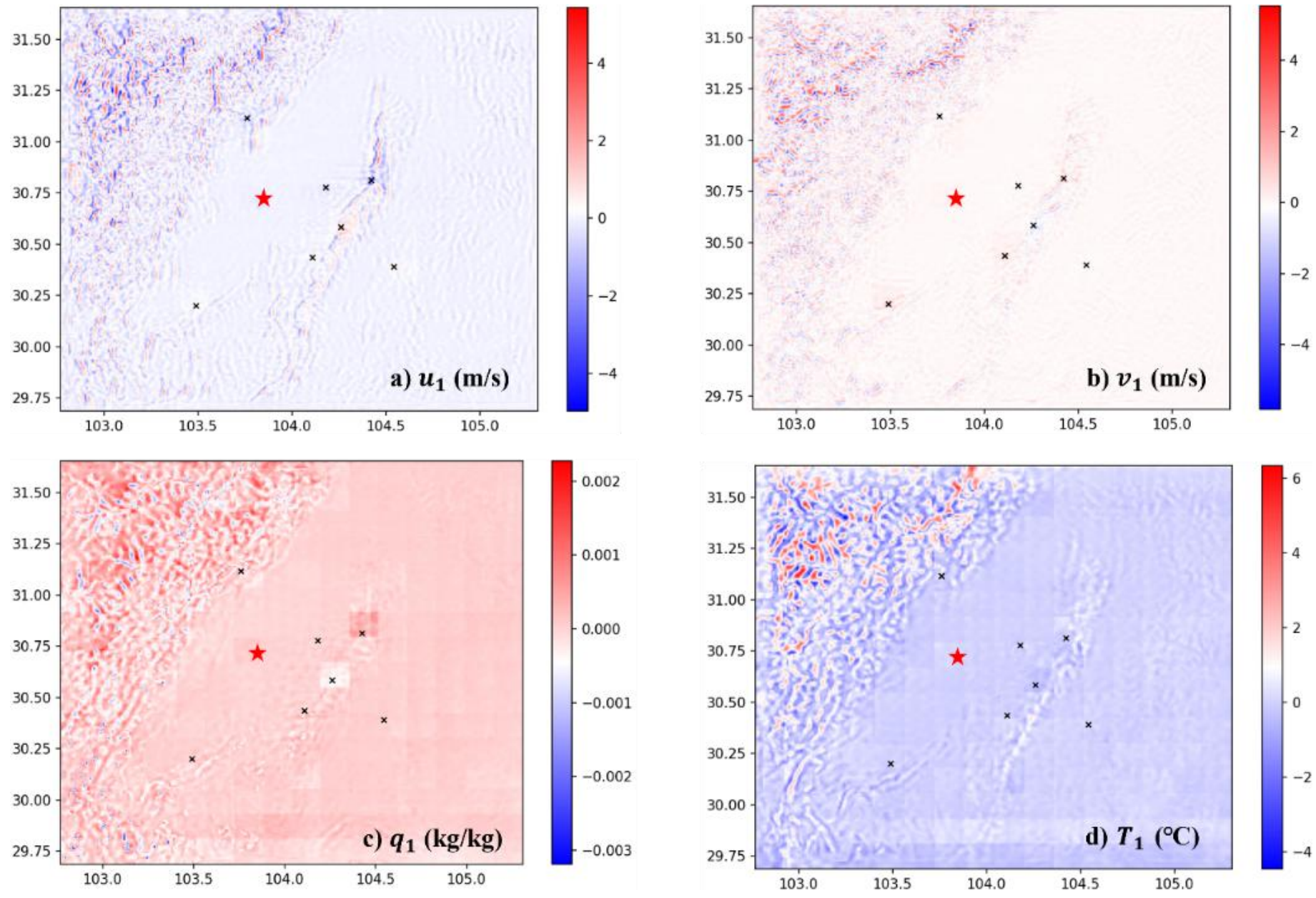

**Figure 12: The distributions of analysis increments (the analysis minus the first guess from background) of the first model level *a*) u-component $u_1$ and *b*) v-component $v_1$ and *c*) temperature $T_1$ as well as *d*) specific humidity $q_1$ at the minimization equilibrium point of the first assimilation step in D03 L3DVar of $H_{f2p}$MDA-$WRF_{FSDA}$. The $H_{f2p}$MDA-$WRF_{FSDA}$ system is run from 2025-06-21 0000Z to 2025-07-21 0000Z with initial and boundary conditions remapped from ERA5 reanalysis. The black asterisks denote the locations of 8 Lidars. The red asterisk denotes the Lidar location that is used for the evaluation of L3DVar FSDA in Fig. 13.**

Then, we present the time series of RMSEs of the first model level (near surface) u-component (called $u_1$) and v-component (called $v_1$) produced by the classic multiscale DA (denoted by MDA) and new AI L3DVar (denoted by LDA) in **Fig. 13**. Within the $H_{f2p}$MDA-$WRF_{DA}$ it is very convenient to perform traditional DA and newly-developed AI L3DVar in a parallel fashion. We see that except for the period of 0600Z-1000Z 21 June 2025, in most of the time, both the MDA and L3DVar reduce the RMSEs from the CTL. As an average in the whole assimilation period, the $u_1$'s RMSE of L3DVar is improved by 2% from the traditional MDA (the RMSEs of CTL, MDA, L3DVar are 2.78, 2.46 and 2.41, respectively). However, we also see that the L3DVar's $v_1$ gives equivalent results of classic MDA (the RMSEs of CTL, MDA, L3DVar are 3.31, 2.92 and 2.93, respectively). Since the model version used in this study does still not resolve turbulence-featured extreme events (1 km horizontal resolution and 100 m vertical resolution in planetary boundary layer in this case), when one of 8 Lidars measures out a local gust event at 06UTC-10UTC 21 June, neither MDA nor L3DVar does not capture this event. This leads to large RMSEs during this period for all of CTL, MDA and L3DVar. Future studies will first detect the sensitivities of these DA schemes on model horizontal and vertical resolutions, and then perform deep research on L3DVar to get optimal results, serving for low-altitude economy environment safety insurance.

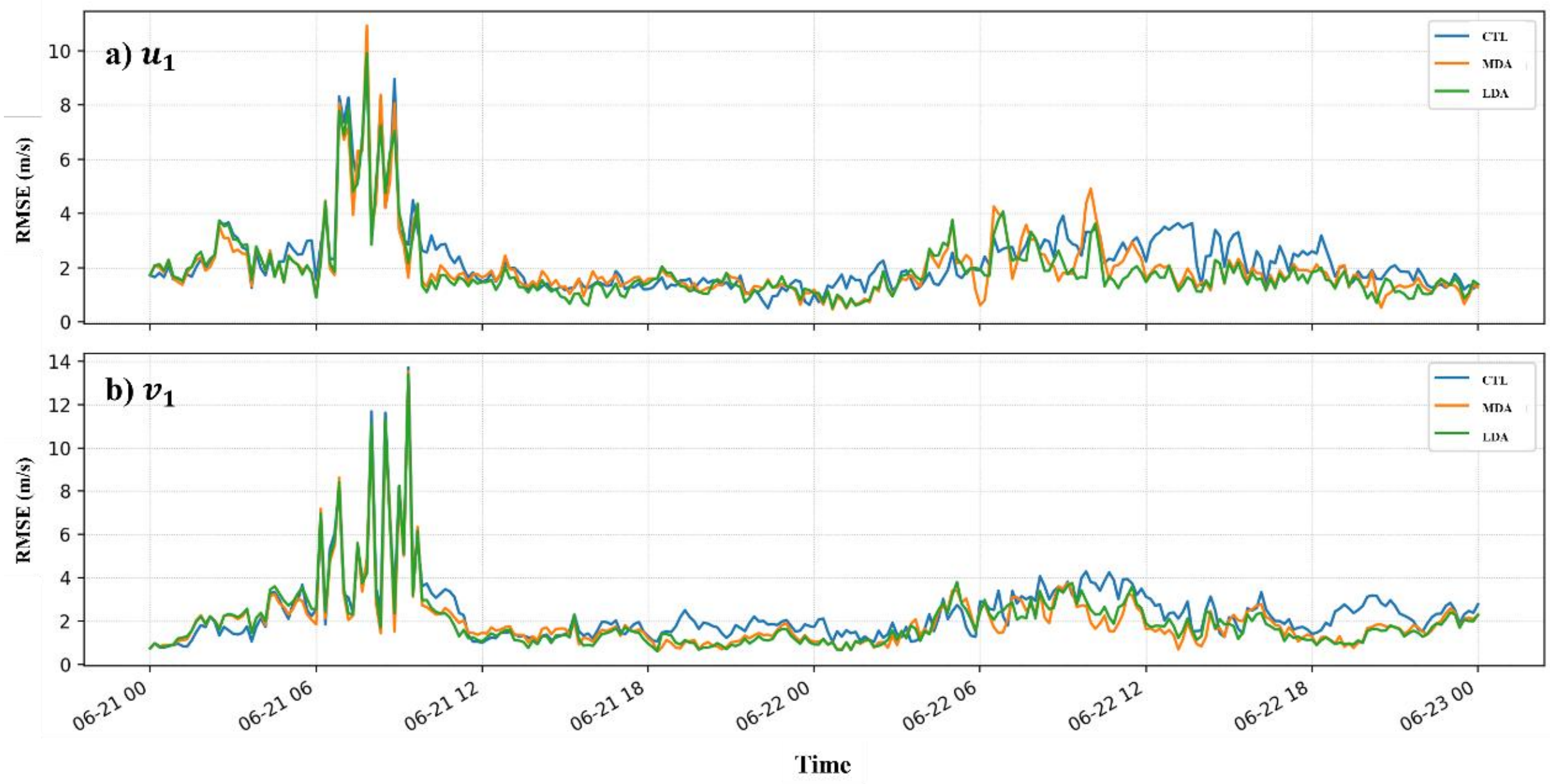

**Figure 13: The time series of root mean squared errors (RMSEs) of the first model level u-component ($u_1$) and v-component $v_1$ in free model control simulation (CTL) (blue), and multiscale high-efficient approximate filter (MSHea-EnKF) data assimilation (MDA, orange) as well as AE latent data assimilation (LDA, green) produced by Hf2pMDA-$WRF_{FSDA}$ in the D03 domain. Both MDA and LDA experiments are conducted under the configuration of multi-layer nested downscaling shown in Fig. 9*a* in which the WRF model with 9/3 (1) km resolutions in D01/D02 (D03) assimilates conventional surface station (Lidar) observations shown in Figs. 9*b*-*c*. What are shown here are the results of the first two days in the one month test period of 2025-06-21 0000Z to 2025-07-21 0000Z for a Lidar location denoted by a red asterisk in Fig. 12.**

We use **Fig. 14** to show that although we only assimilate the surface wind measurements by Lidar into the downscaled model, the online FSDA system is able to propagate the surface observational information vertically to constrain the model. While the surface-only L3DVar makes difference from the classic MDA (**panel *a***), the small surface absolute error can impact on other model levels gradually (**panels *b-g***). However, this surface-only observational constraint is clearly insufficient and follow-up studies shall use the profiles measured by Lidar to gain more vertical observational constraints.

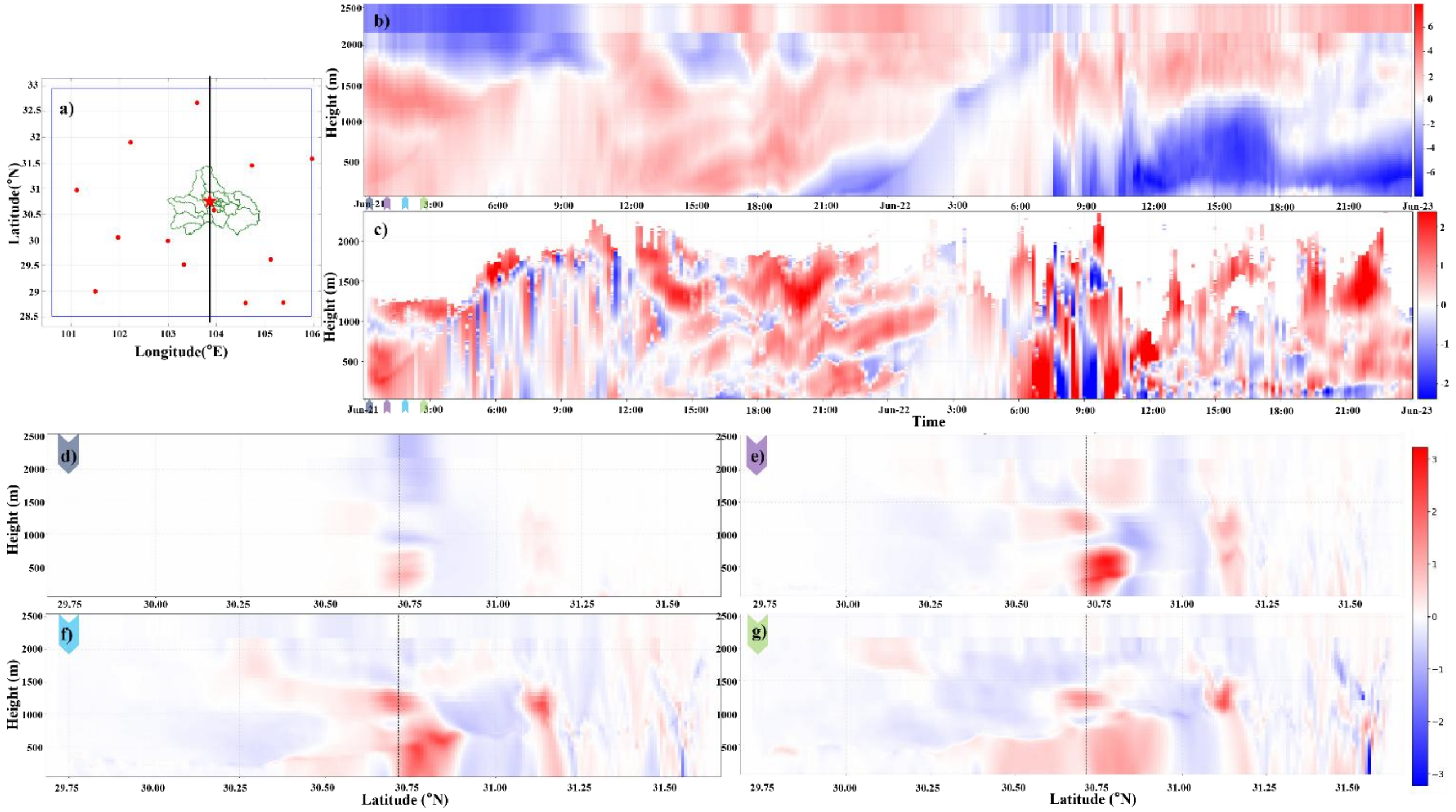

**Figure 14: Vertical structures of atmospheric wind u-component. (a) Locations of the lidar sites. The selected lidar site is marked by the red asterisk (the same Lidar denoted by a red asterisk in Fig. 12), and the black line indicates the radial transect used for the vertical cross-sections in panels (d–g). (b) Time–height distribution of the L3DVar minus MDA difference at the selected Lidar site. (c) Time–height distribution of differences of u-component's absolute error of L3DVar minus MDA. The colored markers below panels (b) and (c) indicate the time snapshots corresponding to the cross-sections in panels (d–g). (d–g) Vertical cross-sections of u-component's difference of L3DVar minus MDA along the radial transect at (d) 0010Z, (e) 0100Z, (f) 0150Z, and (g) 0240Z on 2025-06-21.**

## 6 Summary and discussions

Based on the F2PY protocol, a Fortran-Python hybrid programming infrastructure platform called $H_{f2p}$MDA has been developed for deep incorporation of AI and scientific modeling and data assimilation. In this $H_{f2p}$MDA framework, any Python-coded (Fortran-coded) AI (scientific) algorithm (scheme) can be used by a scientific modeling and data assimilation module (machine learning procedure). As two typical application examples of $H_{f2p}$MDA, the implementations with a coupled global climate model CM2 with modularized Fortran structures and a multi-layer nesting downscaling weather model WRF featured as recursive Fortran integration, have been presented in details. The CM2's (WRF's) $H_{f2p}$MDA implementation establishes a nonlinear strongly-coupled (high-precision) climate (weather) data assimilation system called $H_{f2p}$MDA-$CM2_{SCDA}$ ($H_{f2p}$MDA-$WRF_{FSDA}$). Test results of $H_{f2p}$MDA-$CM2_{SCDA}$ ($H_{f2p}$MDA-$WRF_{FSDA}$) show that while the $H_{f2p}$MDA conveniently inserts an AI latent DA algorithm into the $CM2_{CDA}$ ($WRF_{DA}$) system and fulfills global strongly-CDA (regional high-precision DA), it readily improves the assimilation quality from a traditional DA method. Except for deeply evaluating scientific values in $H_{f2p}$MDA-$CM2_{SCDA}$ and $H_{f2p}$MDA-$WRF_{FSDA}$, follow-up studies also include incorporating more advanced AI DA algorithms into $H_{f2p}$MDA, for example, the generative assimilation and prediction (GAP) (Yang et al., 2025) based on image recognition technology.

More generally, the $H_{f2p}$MDA designed in this study is a highly efficient infrastructure platform for deep incorporation of AI and science since it only requires minimum changes on legacy Fortran codes of sciences that have a long persistently-developing history. Following the procedure of $H_{f2p}$MDA, on the one hand, it is feasible to incorporate data-driven machine deep learning algorithms to improve scientific modeling to fulfill "AI for science." For example, there are urgent demands of new parameterization schemes for high-resolution (HR) Earth system model (e.g. Chang et al., 2020) or even resolving the challenging problems of HR Earth system modeling (e.g. Gou et al., 2025). On the other hand, the rich achievements on scientific modeling from long time development over more than a half century since 1950s can directly advance development of science-guided AI algorithms (e.g. Hao et al., 2025) to advance "science for AI." In that sense, while the $H_{f2p}$MDA has a very wide scope of applications, it requires more efforts to optimize its structure and deepen its infrastructure development so that it can be more conveniently plugged in any application scenario.

**Code and data availability**

The original ERA5 dataset (Hersbach et al., 2020) can be obtained from https://doi.org/10.24381/cds.adbb2d47.

The original OISST v2.1 sea surface temperature dataset (Huang et al., 2021) can be obtained from https://www.ncei.noaa.gov/products/optimum-interpolation-sst.

The CM2.1 model (Delworth et al., 2006b) can be obtained from https://github.com/mom-ocean/MOM5 and the CM2.1 model version with DA modules is also archived on Zenodo (https://doi.org/10.5281/zenodo.18883209; Delworth et al., 2006a).

The Weather Research and Forecasting model version 3.7.1 (WRF v3.7.1; Skamarock et al., 2008) can be obtained from https://www2.mmm.ucar.edu/wrf/users/download/get_source.html. The exact version used in this study has also been archived to ensure reproducibility and long-term accessibility (https://doi.org/10.5281/zenodo.19271007; University Corporation for Atmospheric Research and NSF National Center for Atmospheric Research, 2015).

The model code for $H_{f2p}$MDA-CM2$_{SCDA}$ and $H_{f2p}$MDA-WRF$_{FSDA}$ developed in this study is archived at https://doi.org/10.5281/zenodo.20710848 (Zhu et al., 2026b). The datasets used in the experiments, including the observation data and the exact ERA5 and OISST data used in this study, are archived at https://doi.org/10.5281/zenodo.19272242 (Zhu et al., 2026a).

**Author contributions**

XZ and ZLin are co-first authors who contributed equally to this work. Both of them conduct all test work on Python-Fortran hybrid programming as well as AE and latent minimization and join the paper edits. SZ is the corresponding author who proposes the idea, designs and organizes the research project, writes and edits the paper. ZLu participated the test experiments on WRF multiscale data assimilation and joins the discussions. SW and XH manage the Leice Transient team and OUC team respectively, and join the discussions and make comments on the research work. ZX participated the test experiments on WRF AE training experiments and joins the discussions. ZR, JL, JX, YG, RH, XY, ML and GL contribute to this research by joining the discussions and making comments on the research and manuscript.

**Competing interests**

The authors declare that they have no conflict of interest.

**Acknowledgements**

This work is supported by the Science and Technology Innovation Project of Laoshan Laboratory under Grants (Nos. LSKJ202300400, LSKJ202300401-03, LSKJ202202200, LSKJ202202201-04), the National Natural Science Foundation of China (42361164616), the Shandong Province's "Taishan" Scientist Program (ts201712017).

**Review statement**

The review statement will be added by Copernicus Publications listing the handling editor as well as all contributing referees according to their status anonymous or identified.

# Supplementary Information for

# Python-Fortran Hybrid Programming to Fuse AI and Physical Models: Examples of AI-LDA in climate and weather models (Hf2pMDA_v1.0)

Xianrui Zhu[1+], Zikuan Lin[2+], Shaoqing Zhang[2*], Zebin Lu[2], Songhua Wu[3,4], Xiangyun Hou[2], Zhisheng Xiao[3], Zhicheng Ren[3], Jiangyu Li[5], Jing Xu[5], Yang Gao[6], Rixu Hao[7], Xiaolin Yu[2], Mingkui Li[2], Guangliang Liu[8]

[1]Chongben Honors College, Ocean University of China, Qingdao, 266100, China

[2]Key Laboratory of Physical Oceanography, Ministry of Education, and Institute for Advanced Ocean Study, and Frontiers Science Center for Deep Ocean Multispheres and Earth System (FDOMES), College of Oceanic and Atmospheric Sciences, Ocean University of China, Qingdao, 266100, China

[3]Qingdao Leice Transient Technology Co., Ltd., Qingdao, 266100, China

[4]College of Marine Technology, Ocean Remote Sensing Institute, Ocean University of China, Qingdao, China

[5]Qingdao Marine and Meteorological Institute, Qingdao, 266100, China

[6]Key Laboratory of Marine Environmental Science and Ecology, Ministry of Education, Frontiers Science Center for Deep Ocean Multispheres and Earth System (FDOMES), Ocean University of China, Qingdao, 266100, China

[7]College of Intelligent Systems Science and Engineering, and Engineering Research Center of Navigation Instruments, Ministry of Education, Harbin Engineering University, Harbin, 150001, China.

[8]State Key Laboratory of Physical Oceanography and Artificial Intelligence, Jinan, China.

**Contents of this file:**

**Text S1: General procedure of implementation of $H_{f2p}$MDA**

A general procedure of implementing Python-Fortran hybrid programming for Deep Incorporation of AI and Physical Models and data assimilation based on F2PY ($H_{f2p}$MDA**)** includes the following 4 steps.

**Step-1: Environment setting**

In theory, there is no specific environment variable for F2PY. However, to minimize the interruption of other computation tasks to the Python-Fortran modeling development, we strongly recommend to use an environment setting tool such as *conda* etc. to set an isolated environment for the Python-Fortran hybrid application. We may start from a Python environment by typing the command "*conda create -n f2py python=3.11*" and continue to complete the setting by typing "*conda activate f2py*" and "*conda install numpy=1.26*" for follow-ups until necessary, where 3.11 and 1.26 are environment parameters of corresponding software. For some non-standardized environment variables only necessary for specific applications, the "*source*" command is a convenient complementary. Detailed description will be given in specific applications by **Sects. 3 and 4**.

Here is an example of setting environment for *openmpi* which is an important application of MPI in our $H_{f2p}$MDA-CM2$_{CDA}$ and $H_{f2p}$MDA-WRF$_{DA}$ cases. We need to recompile *openmpi* package with added "-fPIC" tag and "*make install*". Then we *source* a file of command list containing the following 3 lines to add the *openmpi* application to the environment *myf2p*:

```
export PATH=".../openmpi/bin:$PATH"
export LIBRARY_PATH=".../openmpi/lib:$LIBRARY_PATH"
export LD_LIBRARY_PATH=".../openmpi/lib:$LD_LIBRARY_PATH"
```

Again, the *openmpi* directory shall be in the specific application directory for $H_{f2p}$MDA-CM2$_{CDA}$ so that this *openmpi* setting does NOT influence any other applications.

**Step-2: Creating of Python-callable Fortran (PCF) modules**

In order to become Python-callable, the existed Fortran codes need to be recompiled with an additional new tag "-fPIC" that makes the Fortran codes being compiled become Dynamic Link Library (DLL) candidates with a "position independent" nature. Once relatively-independent Fortran codes are all recompiled as DLL candidates, they are ready to link with the Python main controller (PMC).

To minimize complexity of the infrastructure between PMC and PCF modules, a plug-featured Fortran subroutine that well organizes the PCF modules and makes a few plug-ins for PMC needs to be encapsulated with all PCF modules to form a signature file that shows a clear structure of such plugs. We use the following command line to initially create a signature file called *appname.pyf*:

```
$ f2py app_plugs.F90 -m appname -h --overwrite-signature appname.pyf
```

where *app_plugs.F90* consists of subroutines or functions that organize the PCF modules and serve as the plug-ins of PMC. Supposing that there are a few DLL candidates created in the compiling procedure described above called *libapp1.a*, *libapp2.a,…*etc., then, the encapsulating command looks like a continuous line of the above initial creation line of *appname.pyf* as:

```
$ FC="mpif90" CC="mpicc" CXX="mpicxx" LDSHARED="mpif90" LDFLAGS="-no-ipo \
-Wl, --export-dynamic" f2py -c appname.pyf app_plugs.F90 libapp1.a \
libapp2.a ... -lnetcdf -L/.../software/netcdf3/lib -L/.../openmpi/lib \
--backend distutils
```

where the option "-no-ipo" closes out Interprocedural Optimization to minimize potential uncertainties due to optimization cross modules and the option "-Wl" ensures used functions linked to DLL. Note that here the openmpi/lib also needs to be recompiled with the tag "-fPIC." Finally, as the above command is normally conducted without any error message, a complete DLL file called *appname.cpython-311-x86_64-linux-gnu.so* is formed (automatically named by the compiling environment parameters), which is a Python-callable DLL "shared object" where all subroutines and functions are callable by PMC. All information about the logic structure of *appname.cpython-311-x86_64-linux-gnu.so* is recorded in the signature file *appname.pyf*, which will be served as an import name in PMC, also serving as an efficient guideline for constructing PMC.

**Step-3: Creating of Fortran-callable Python (FCP) modules**

To let the Fortran codes be able to call Python-coded algorithms, we need an interface (called *callback_python.F90*, for instance) that translates the Python external functions into Fortran-callable public subroutines (called *python_algorithm{1,2}* etc., for instance). At the same time, the *app_plug.F90* shall include the definition of these external functions (called *python_foo{1,2}* etc., for instance) which are transported from PMC into Fortran application at plug-ins in *app_plug.F90* as arguments. This Fortran interface file *callback_python.F90* shall be compiled with *app_plug.F90* together by going through the procedure described in Step 2 to join the signature file *appname.pyf*. Then, all subroutines that represent python algorithms can be called in any Fortran application through a normal "*use*" statement.

**Step-4: Creating and executing of Python main control (PMC) program**

A PMC (called *appname_main.py*, for instance) shall consist of three important parts: import statements, model integration conductor and AI algorithm interface as described before. The import statements must include "from *appname* import *module_list*" (import selected Fortran modules from *appname* defined before) or simply "import *appname*" to import all modules in *appname*. The import statements may include any necessary modules that are used in PMC such as an MPI module which provides process-element identification (PE-id) etc. information, as well as defined AI algorithms and so on.

To minimize Python-recoding for the existed Fortran model, the PMC's statements for conducting model integration usually remain as simple as possible Python AI algorithms are conveniently called by the Fortran model. In the example of climate model application of $H_{f2p}$MDA, which will be described in **Sect. 4**, we implement an AI strongly-coupled data assimilation (SCDA) based on the existed Fortran-coded weakly-coupled data assimilation (WCDA). Due to the strongly-modular nature of existed Fortran codes, the AI algorithm trained by deep-learning is conveniently inserted at the atmosphere-ocean interface, and we let PMC to take care of time integration loop (detailed in **Sect. 4.2**). However, in the example of $H_{f2p}$MDA's NWP model application described in **Sect. 5**, we implement AI DA in the 3$^{rd}$ domain downscaled from a recursive domain integration. In that circumstance, the PMC's model integration conductor only consists of a single line, which will be elucidated in **Sect. 5.2**.

Once a PMC is ready, we can still use *mpirun* command to reflect the MPI nature of Python-Fortran hybrid computation as:

```
$ mpirun -n xx python appname_main.py > appname.log
```

where "xx" is the PE number required by the program, *appname.log* is the log file recording the log information during the program running. We may insert this run command line into the existed *run_script* file of the Fortran model to let the $H_{f2p}$MDA run in a large scale of MPI usage by a *qsub* command to submit it to the back-end job queue.

## Text S2: Implementation details of an AE architecture

### S2.1 General architecture design of the AE for SCDA in $H_{f2p}$MDA-CM2$_{SCDA}$

The autoencoder (AE) used for SCDA in $H_{f2p}$MDA-CM2$_{SCDA}$ is illustrated in **Fig. S1**. The model adopts a two-branch architecture because the atmospheric and oceanic input fields have different spatial resolutions. In the encoder, the two groups of inputs are first compressed to comparable spatial scales. Their representations are then coupled through two cross-attention stages while the feature dimensions are further compressed. Since the atmospheric and oceanic information has already been exchanged and embedded in the latent representation, the decoder does not include additional cross-branch interaction. Instead, it decodes the latent representation separately for the atmospheric and oceanic output fields. The model contains 10,329,868 trainable parameters, including 1,998,340 parameters in the encoder and 8,331,528 parameters in the decoder.

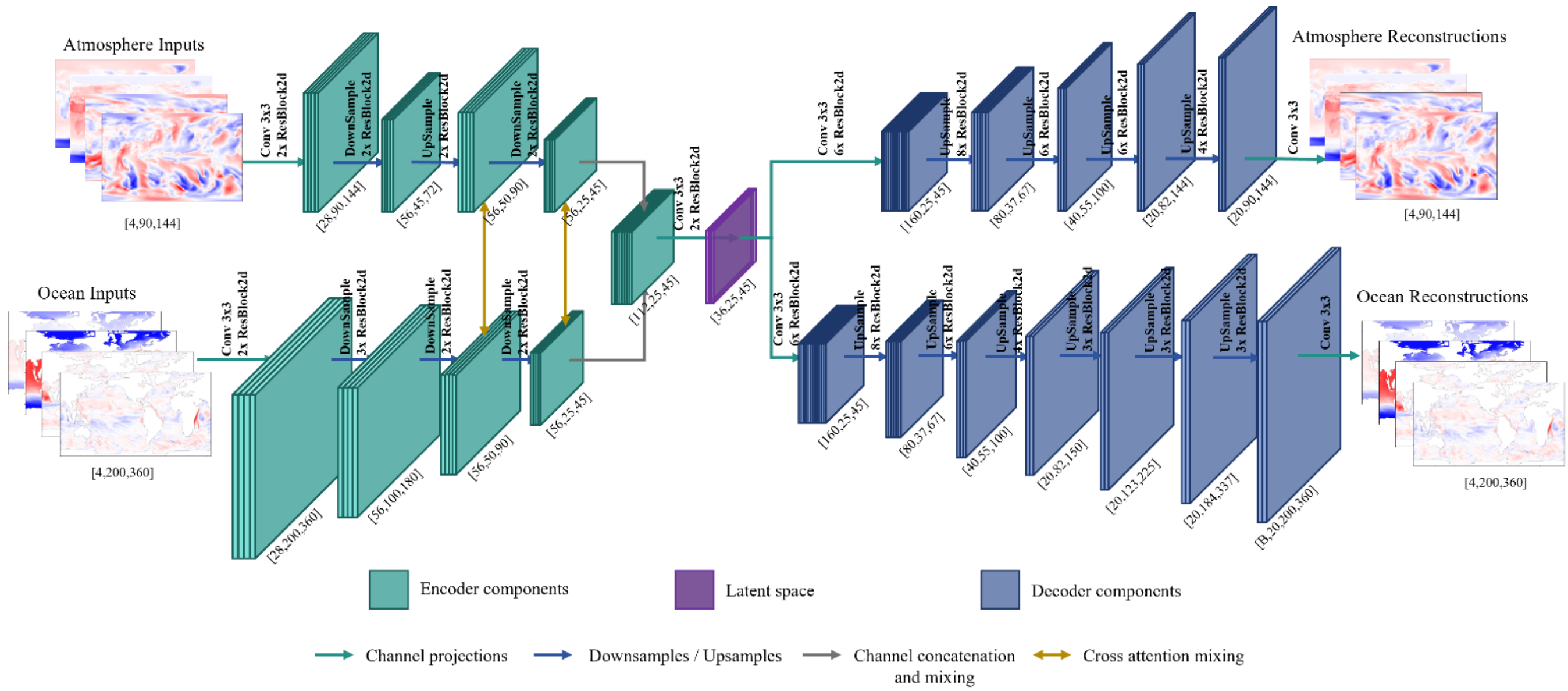

**Figure S1. Architecture of the AE for SCDA in $H_{f2p}$MDA-CM2$_{SCDA}$. The numbers in square brackets indicate the tensor shape after each operation, ordered as channel, height, and width. The labels above the arrows denote the transition modules, which are described in Text S2.1.**

The AE uses standard residual blocks (ResBlocks) to increase network depth and improve representation capacity (He et al., 2016). The downsampling and upsampling stages are implemented using two-dimensional convolutional layers with $3 \times 3$ kernels. In the upsampling stages, interpolation is first applied to increase the spatial resolution, followed by a two-dimensional convolution.

The cross-attention modules follow the scaled dot-product attention formulation introduced in the Transformer architecture (Vaswani et al., 2017). In general, attention is defined as

$$\text{Attention}(Q, K, V) = \text{softmax}\left(\frac{QK^{\top}}{\sqrt{d_k}}\right)V \qquad (S1)$$

where $Q$, $K$, $V$ are the query, key, and value projections, respectively, and $d_k$ is the dimension of the key projection. When the query is derived from one branch and the key and value are derived from another branch, the operation becomes cross attention. In each cross-attention stage, the atmospheric and oceanic branches are updated as

$$\text{Attention}\left(X_a W_Q^a, X_o W_K^o, X_o W_V^o\right)$$
$$\text{Attention}\left(X_o W_Q^o, X_a W_K^a, X_a W_V^a\right) \qquad (S2)$$

where $X_a$ and $X_o$ denote the atmospheric-branch and oceanic-branch features before cross attention, respectively. $W_Q$, $W_K$, and $W_V$ are learnable projection matrices. Through these two directional cross-attention operations, information is exchanged between the atmospheric and oceanic representations before being compressed into the latent space.

### S2.2 Training details of the AE for SCDA in $H_{f2p}$MDA-CM2$_{SCDA}$

#### a) Computing environment and hyperparameters details

The AE for SCDA in $H_{f2p}$MDA-CM2$_{SCDA}$ was trained on four NVIDIA GeForce RTX 4090 GPUs using DeepSpeed (Rasley et al., 2020) and distributed data parallel training. The manually selected checkpoint was taken from epoch 72. The dataset contained 160,600 samples, including 144,540 training samples and 16,060 validation samples. The batch size on each GPU was 4, and 8 gradient-accumulation steps were used, giving an effective batch size of 128.

The model was optimized with AdamW (Loshchilov and Hutter, 2017). The learning rate was $7.5 \times 10^{-5}$, the weight decay was 0.01, and the AdamW parameters were $\beta = (0.9, 0.999)$ and $\varepsilon = 10^{-8}$. Mixed-precision FP16 (Micikevicius et al., 2017) training was used, with gradient clipping set to 1.0. The learning rate followed a cosine schedule with 5 warm-up epochs and a minimum learning rate of $10^{-6}$. The dataloader used 8 workers, a prefetch factor of 32, persistent workers, pinned memory, block shuffling, and drop-last batching.

#### b) Loss details

Before training, the CM2 variables were normalized using z-score normalization:

$$\frac{x-\mu}{\sigma} \tag{S3}$$

where $\mu$ and $\sigma$ are the mean and standard deviation of each variable, respectively. Invalid values were replaced by zero after normalization. Masks were used in the reconstruction loss to exclude invalid grid points where applicable.

The total CM2 loss was formulated as

$$L_{\text{total}} = \eta_{\text{atm}} L_{\text{atm}} + \eta_{\text{ocn}} L_{\text{ocn}} + \lambda_{\text{corrupt}} L_{\text{corrupt}} \tag{S4}$$

where $L_{\text{atm}}$ and $L_{\text{ocn}}$ are the atmospheric and oceanic reconstruction losses, respectively. The domain factors were $\eta_{\text{atm}} = 1.0$ and $\eta_{\text{ocn}} = 1.15$. The latent-corruption weight was $\lambda_{\text{corrupt}} = 0.12$.

For each domain, the loss was defined as

$$L_k = \lambda_1 L_1 + \lambda_2 L_2 + \lambda_g L_{\text{grad}} + \lambda_e L_{\text{energy}} + \lambda_s L_{\text{spectral}} + \lambda_b L_{\text{subband}} \tag{S5}$$

Here, $L_1$ and $L_2$ are the masked absolute and squared reconstruction errors, respectively. $L_{\text{grad}}$ compares horizontal and vertical finite-difference gradients, $L_{\text{energy}}$ compares the spatial mean of squared amplitudes, $L_{\text{spectral}}$ compares log-amplitude Fourier spectra, and $L_{\text{subband}}$ constrains reconstruction errors in different spatial-frequency bands. The main loss weights were

$$\lambda_1 = 1.2, \quad \lambda_2 = 0.35, \quad \lambda_g = 0.10, \quad \lambda_e = 0.05, \quad \lambda_s = 0.008, \quad \lambda_b = 0.45$$

The latent-corruption term used single-pass denoising with Gaussian perturbations applied to the latent representation:

$$\tilde{z} = z + \xi, \qquad \xi \sim \mathcal{N}(0, \sigma_z^2 I), \qquad \sigma_z = 0.012$$

The subband loss was introduced to reduce reconstruction errors across multiple spatial-frequency ranges. For each predicted field $\hat{x}$ and target field $x$, the two-dimensional Fourier transforms are

$$F = \mathcal{F}(x) \quad (S6)$$

The normalized wavenumber radius was divided into four bands:

$$[[0,0.25), \quad [0.25,0.50), \quad [0.50,0.72), \quad [0.72,1.01]]$$

For each band $b$, a frequency mask $M_b$ was applied, and the corresponding band-limited fields were transformed back to physical space:

$$x_b = \mathcal{F}^{-1}(FM_b) \quad (S7)$$

The reconstruction error in each band was measured using a relative Charbonnier penalty (Charbonnier et al., 1997):

$$E_b = \frac{\mathrm{mean}\left[\sqrt{(\hat{x}_b - x_b)^2 + \epsilon^2}\right]}{\max(\mathrm{mean}(|x_b|), \epsilon)} \quad (S8)$$

where $\epsilon = 5 \times 10^{-3}$. The final subband loss was calculated as the weighted mean over all frequency bands:

$$L_{\mathrm{subband}} = \frac{\sum_b w_b E_b}{\sum_b w_b} \quad (S9)$$

In the CM2 configuration, the raw band weights were

$$[0.55,\ 0.85,\ 1.50,\ 2.55]$$

with stronger weights assigned to higher-frequency bands. The implementation further applied a mild frequency-dependent reweighting with power 0.35, yielding effective weights of approximately

$$[0.356,\ 0.808,\ 1.691,\ 3.249]$$

This design is conceptually related to frequency-domain reconstruction losses, which have been used to reduce spectral discrepancies between reconstructed and target images or fields, such as the Focal Frequency Loss of Jiang et al. (2021). The use of the Charbonnier penalty provides a robust and differentiable alternative to the absolute error.

### S2.3 General architecture design of the AE for FSDA in $H_{f2p}$MDA-WRF$_{FSDA}$

The AE used for WRF's fine-scale convection data assimilation (FSDA) in $H_{f2p}$MDA-WRF$_{FSDA}$ is shown in **Fig. S2**. Compared with the SCDA AE, this architecture is more compact and is mainly built from Swin Transformer blocks (Liu et al., 2021). The model contains 53,854,780 trainable parameters, including 41,403,486 parameters in the encoder and 12,451,294 parameters in the decoder.

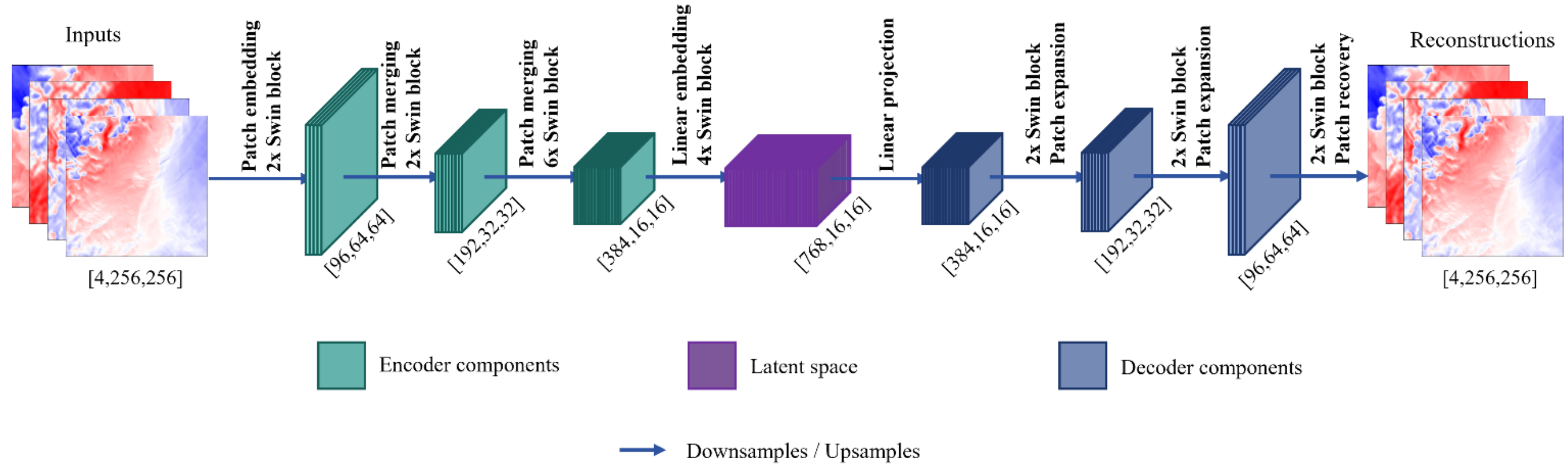

**Figure S2: Architecture of the AE for FSDA in $H_{f2p}$MDA-WRF$_{FSDA}$. The notation follows Fig. S1. The transition modules are described in Text S2.3.**

In **Fig. S2**, the operations are grouped according to the actual workflow of the encoder, latent modeling module, and decoder. Patch embedding denotes the initial strided convolution, which converts gridded input fields into patch tokens and performs the first spatial downsampling. In the encoder, each subsequent transition consists of patch merging followed by Swin Transformer blocks. Patch merging reduces the spatial resolution while increasing the feature dimension.

The final encoder transition uses a linear embedding layer to project the deepest encoder feature into the latent feature dimension, followed by latent Swin Transformer blocks. In the decoder, the first transition is a linear projection that maps the latent feature back to the decoder feature dimension. The following decoder transitions use Swin Transformer blocks together with patch expansion to recover the spatial resolution. The final transition uses patch recovery to reconstruct the output fields. The Swin Transformer blocks follow the shifted-window attention design of Liu et al. (2021), in which self-attention is computed within local windows and cross-window communication is achieved by shifted window partitions. Learnable position embeddings and relative position bias are used inside the Transformer modules, but they are not shown separately in **Fig. S2** for clarity.

### S2.4 Training details of the AE for FSDA in $H_{f2p}$MDA-WRF$_{FSDA}$

#### a) Computing environment and hyperparameters details

The AE for FSDA in $H_{f2p}$MDA-WRF$_{FSDA}$ was trained on four NVIDIA L40 GPUs using PyTorch distributed data parallel training (Li et al., 2020). The dataset contained 26,278 samples and was split by quarter into 21,887 training samples, 2,208 validation samples, and 2,183 test samples. Quarters 1–10 were used for training, quarter 11 for validation, and quarter 12 for testing. The batch size was 256, the maximum number of epochs was 200, and the optimizer was Adam (Kingma and Ba, 2014) with a learning rate of $10^{-4}$. The learning rate scheduler was ReduceLROnPlateau with `mode="min"`, `factor=0.5`, and `patience=10`. The dataloader used 32 workers and pinned memory.

**b) Loss details**

The WRF variables U10, V10, T2, and Q2 were normalized using fixed z-score statistics that are defined by **eq. S3**. Missing values were replaced by zero, and all fields were bilinearly interpolated to $256 \times 256$. The training target was the input field itself; therefore, the task was deterministic autoencoding.

The FSDA AE was trained with a combination of mean squared error (MSE) and structural similarity loss. The structural similarity index measure (SSIM) was originally proposed by Wang et al. (2004). For two fields $x$ and $y$, SSIM is defined as

$$\mathrm{SSIM}(\mathrm{x},\mathrm{y}) = \frac{(2\mu_x\mu_y + C_1)(2\sigma_{xy} + C_2)}{(\mu_x^2 + \mu_y^2 + C_1)(\sigma_x^2 + \sigma_y^2 + C_2)} \quad (S10)$$

where $\mu_x$ and $\mu_y$ are local means, $\sigma_x^2$ and $\sigma_y^2$ are local variances, $\sigma_{xy}$ is the local covariance, and $C_1$ and $C_2$ are small constants used for numerical stability.

The final training loss was

$$L = \mathrm{MSE}(\hat{y} \odot M, y \odot M) + 0.2[1 - \mathrm{SSIM}(\hat{y} \odot M, y \odot M)] \quad (S11)$$

where $\hat{\boldsymbol{y}}$ is the reconstructed field, $\boldsymbol{y}$ is the target field, $\boldsymbol{M}$ is the mask, and $\odot$ denotes element-wise multiplication. The MSE term constrains pointwise reconstruction accuracy, whereas the SSIM term provides an additional structural constraint on the reconstructed spatial pattern.

**Text S3: L3DVar running details**

This section summarizes the optimization settings and computational cost of L3DVar in the two assimilation applications. The weighted objective function used in both configurations is defined as

$$J(\mathbf{z}) = \frac{1}{2}\|\mathbf{z} - \mathbf{z}_b\|^2_{\mathbf{B}_z^{-1}} + \frac{\lambda}{2}\left\|\mathbf{y} - \mathcal{H}\big(D(\mathbf{z})\big)\right\|^2_{\mathbf{R}^{-1}} \quad (S12)$$

where $\lambda$ controls the relative contribution of the observation term. The estimation of $\mathbf{B}_z$ for the two applications is described in **Sect. 4.3.2 and 5.3.2**. The corresponding covariance settings and optimization parameters are listed in **Table S1**.

**Table S1. L3DVar optimization settings.**

| Application | $\lambda$ | Number of iterations | Learning rate | Optimizer |
|---|---|---|---|---|
| SCDA by $H_{f2p}$MDA-CM2$_{SCDA}$ | 100 | 100 | $10^{-3}$ | Adam |
| FSDA by $H_{f2p}$MDA-WRF$_{FSDA}$ | 1000 | 300 | $10^{-3}$ | Adam |

In addition to the optimization configuration, the computational contribution of L3DVar was evaluated for the FSDA by $H_{f2p}$MDA-WRF$_{FSDA}$ experiment. **Table S2** reports the accumulated DA cost relative to the main model runtime and the total runtime including output writing.

**Table S2. Computational efficiency of L3DVar in the SCDA by $H_{f2p}$MDA-CM2$_{SCDA}$ and FSDA by $H_{f2p}$MDA-WRF$_{FSDA}$ experiments. DA total denotes the accumulated runtime of the L3DVar optimization, Total main denotes the total runtime of the main workflow, and DA/main is their ratio. The f2py overhead denotes the additional runtime introduced by the Python–Fortran interface, and Interface efficiency denotes the fraction of the DA runtime excluding this overhead.**

| Case | DA total | Total main | DA/main | f2py overhead | Interface efficiency |
|---|---|---|---|---|---|
| SCDA by $H_{f2p}$MDA-CM2$_{SCDA}$ | 3052.7927 s | 22740.1087 s | 13.42% | - | - |
| FSDA by $H_{f2p}$MDA-WRF$_{FSDA}$ | 2782.3042 s | 28053.3913 s | 9.92% | 0.3129 s | 99.99% |

L3DVar accounted for 13.42% and 9.92% of the total main runtime in the SCDA by $H_{f2p}$MDA-CM2$_{SCDA}$ and FSDA by $H_{f2p}$MDA-WRF$_{FSDA}$ experiments, respectively. The f2py overhead was negligible in the latter experiment. The interface overhead was not measured for SCDA by $H_{f2p}$MDA-CM2$_{SCDA}$.

## References S1